\pdfoutput=1
\documentclass[12pt,a4paper]{article}

\usepackage{ifthen} 
\newboolean{pdflatex}
\setboolean{pdflatex}{true} 

\newboolean{articletitles}
\setboolean{articletitles}{true} 

\newboolean{uprightparticles}
\setboolean{uprightparticles}{false} 

\def\paperauthors{LHCb collaboration} 
\def\paperasciititle{Observation of the very rare Sigma+ -> p mu+ mu- decay} 
\def\papertitle{Observation of the very rare \sigmapmumu decay} 
\def\paperkeywords{{High Energy Physics}, {LHCb}, {Rare decay}, {Branching fraction}} 
\def\papercopyright{\the\year\ CERN for the benefit of the LHCb collaboration} 
\def\paperlicence{CC BY 4.0 licence}
\def\paperlicenceurl{https://creativecommons.org/licenses/by/4.0/}

\newif\ifEnableSectionTOCLinks
\EnableSectionTOCLinksfalse 

\usepackage[top=1in, bottom=1.25in, left=1in, right=1in]{geometry}

\usepackage{microtype}
\usepackage{lineno}  
\usepackage{xspace} 
\usepackage{caption} 

\usepackage{graphicx}  
\usepackage{color}
\usepackage{colortbl}
\graphicspath{{./figs/}} 

\usepackage{amsmath} 
\usepackage{amssymb}
\usepackage{amsfonts}
\usepackage{upgreek} 

\newcommand*\patchAmsMathEnvironmentForLineno[1]{%
\expandafter\let\csname old#1\expandafter\endcsname\csname #1\endcsname
\expandafter\let\csname oldend#1\expandafter\endcsname\csname
end#1\endcsname
 \renewenvironment{#1}%
   {\linenomath\csname old#1\endcsname}%
   {\csname oldend#1\endcsname\endlinenomath}%
}
\newcommand*\patchBothAmsMathEnvironmentsForLineno[1]{%
  \patchAmsMathEnvironmentForLineno{#1}%
  \patchAmsMathEnvironmentForLineno{#1*}%
}
\AtBeginDocument{%
\patchBothAmsMathEnvironmentsForLineno{equation}%
\patchBothAmsMathEnvironmentsForLineno{align}%
\patchBothAmsMathEnvironmentsForLineno{flalign}%
\patchBothAmsMathEnvironmentsForLineno{alignat}%
\patchBothAmsMathEnvironmentsForLineno{gather}%
\patchBothAmsMathEnvironmentsForLineno{multline}%
\patchBothAmsMathEnvironmentsForLineno{eqnarray}%
}

\usepackage{hyperxmp}

\usepackage[pdftex,
            pdfauthor={\paperauthors},
            pdftitle={\paperasciititle},
            pdfkeywords={\paperkeywords},
            pdfcopyright={Copyright (C) \papercopyright},
            pdflicenseurl={\paperlicenceurl}]{hyperref}

\usepackage[bottom,flushmargin,hang,multiple]{footmisc}

\usepackage[all]{hypcap} 

\usepackage{xspace} 
\usepackage{upgreek}

\def\MagUp {\mbox{\em Mag\kern -0.05em Up}\xspace}

\ifthenelse{\boolean{uprightparticles}}%
{

 \def\Ppi         {\ensuremath{\uppi}\xspace}

 \def\PDelta      {\ensuremath{\Delta}\xspace}                 
 \def\PXi         {\ensuremath{\Xi}\xspace}                 
 \def\PLambda     {\ensuremath{\Lambda}\xspace}                 
 \def\PSigma      {\ensuremath{\Sigma}\xspace}                 
 \def\POmega      {\ensuremath{\Omega}\xspace}                 
 \def\PUpsilon    {\ensuremath{\Upsilon}\xspace}
 \let\oldPi\Pi
 \def\PPi         {\ensuremath{\oldPi}\xspace}

 \def\PB      {\ensuremath{\mathrm{B}}\xspace}                 
 \def\PD      {\ensuremath{\mathrm{D}}\xspace}                 

 \def\PK      {\ensuremath{\mathrm{K}}\xspace}                 
 \def\Pp      {\ensuremath{\mathrm{p}}\xspace}                 

 \def\Ps      {\ensuremath{\mathrm{s}}\xspace}

 \def\thebaroffset{0.0em}
}
{

 \def\Ppi         {\ensuremath{\pi}\xspace}

 \mathchardef\PDelta="7101
 \mathchardef\PXi="7104
 \mathchardef\PLambda="7103
 \mathchardef\PSigma="7106
 \mathchardef\POmega="710A
 \mathchardef\PUpsilon="7107
 \mathchardef\PPi="7105
 \def\PB      {\ensuremath{B}\xspace}                 
 \def\PD      {\ensuremath{D}\xspace}                 

 \def\PK      {\ensuremath{K}\xspace}                 
 \def\Pp      {\ensuremath{p}\xspace}                 

 \def\Ps      {\ensuremath{s}\xspace}

 \def\thebaroffset{0.18em}
}
\newcommand{\offsetoverline}[2][\thebaroffset]{\kern #1\overline{\kern -#1 #2}}%

\makeatletter
\ifcase \@ptsize \relax
  \newcommand{\miniscule}{\@setfontsize\miniscule{4}{5}}
\or
  \newcommand{\miniscule}{\@setfontsize\miniscule{5}{6}}
\or
  \newcommand{\miniscule}{\@setfontsize\miniscule{5}{6}}
\fi
\makeatother

\DeclareRobustCommand{\optbar}[1]{\shortstack{{\miniscule (\rule[.5ex]{1.25em}{.18mm})}
  \\ [-.7ex] $#1$}}

\def\squark    {{\ensuremath{\Ps}}\xspace}

\def\kaon    {{\ensuremath{\PK}}\xspace}

\def\KorKbar {\kern \thebaroffset\optbar{\kern -\thebaroffset \PK}{}\xspace}

\def\KS      {{\ensuremath{\kaon^0_{\mathrm{S}}}}\xspace}

\def\D       {{\ensuremath{\PD}}\xspace}

\def\DorDbar {\kern \thebaroffset\optbar{\kern -\thebaroffset \PD}\xspace}

\def\Dp      {{\ensuremath{\D^+}}\xspace}
\def\Dm      {{\ensuremath{\D^-}}\xspace}

\def\DpDm    {\ensuremath{\Dp {\kern -0.16em \Dm}}\xspace}

\def\B       {{\ensuremath{\PB}}\xspace}

\def\BorBbar {\kern \thebaroffset\optbar{\kern -\thebaroffset \PB}\xspace}

\def\Bd      {{\ensuremath{\B^0}}\xspace}

\def\BdorBdbar {\kern \thebaroffset\optbar{\kern -\thebaroffset \Bd}\xspace}

\def\Bs      {{\ensuremath{\B^0_\squark}}\xspace}

\def\BsorBsbar {\kern \thebaroffset\optbar{\kern -\thebaroffset \Bs}\xspace}

\def\Y#1S{\ensuremath{\PUpsilon{(#1S)}}\xspace}

\def\proton      {{\ensuremath{\Pp}}\xspace}
\def\antiproton  {{\ensuremath{\overline \proton}}\xspace}

\def\Lz          {{\ensuremath{\PLambda}}\xspace}

\def\LorLbar     {\kern \thebaroffset\optbar{\kern -\thebaroffset \PLambda}\xspace}

\def\Sigmares    {{\ensuremath{\PSigma}}\xspace}

\def\Sigmap      {{\ensuremath{\Sigmares{}^+}}\xspace}

\def\Sigmaresbar {{\ensuremath{\offsetoverline{\Sigmares}}}\xspace}

\def\Sigmabarm   {{\ensuremath{\Sigmaresbar{}^-}}\xspace}

\newcommand{\decay}[2]{\mbox{\ensuremath{#1\!\to #2}}\xspace} 

\def\to                 {\ensuremath{\rightarrow}\xspace}

\def\AT#1     {\ensuremath{A_{\mathrm{T}}^{#1}}\xspace}           

\def\C#1      {\ensuremath{\mathcal{C}_{#1}}\xspace}                       
\def\Cp#1     {\ensuremath{\mathcal{C}_{#1}^{'}}\xspace}                    
\def\Ceff#1   {\ensuremath{\mathcal{C}_{#1}^{\mathrm{(eff)}}}\xspace}        
\def\Cpeff#1  {\ensuremath{\mathcal{C}_{#1}^{'\mathrm{(eff)}}}\xspace}       
\def\Ope#1    {\ensuremath{\mathcal{O}_{#1}}\xspace}                       
\def\Opep#1   {\ensuremath{\mathcal{O}_{#1}^{'}}\xspace}                    

\newcommand{\aunit}[1]{\ensuremath{\text{\,#1}}}       

\newcommand{\tev}{\aunit{Te\kern -0.1em V}\xspace}
\newcommand{\gev}{\aunit{Ge\kern -0.1em V}\xspace}
\newcommand{\mev}{\aunit{Me\kern -0.1em V}\xspace}
\newcommand{\kev}{\aunit{ke\kern -0.1em V}\xspace}
\newcommand{\ev}{\aunit{e\kern -0.1em V}\xspace}
 
\newcommand{\mevc}{\ensuremath{\aunit{Me\kern -0.1em V\!/}c}\xspace}
\newcommand{\gevc}{\ensuremath{\aunit{Ge\kern -0.1em V\!/}c}\xspace}
\newcommand{\mevcc}{\ensuremath{\aunit{Me\kern -0.1em V\!/}c^2}\xspace}
\newcommand{\gevcc}{\ensuremath{\aunit{Ge\kern -0.1em V\!/}c^2}\xspace}

\def\fb   {\ensuremath{\aunit{fb}}\xspace}
\def\invfb   {\ensuremath{\fb^{-1}}\xspace}

\def\sec  {\ensuremath{\aunit{s}}\xspace}

\def\gsim{{~\raise.15em\hbox{$>$}\kern-.85em
          \lower.35em\hbox{$\sim$}~}\xspace}
\def\lsim{{~\raise.15em\hbox{$<$}\kern-.85em
          \lower.35em\hbox{$\sim$}~}\xspace}

\def\sqs   {\ensuremath{\protect\sqrt{s}}\xspace}

\def\pt         {\ensuremath{p_{\mathrm{T}}}\xspace}

\def\tell1  {TELL1\xspace}
\def\ukl1   {UKL1\xspace}

\newcommand{\ie}{\mbox{\itshape i.e.}\xspace}

\newcommand{\lhcborcid}[1]{\href{https://orcid.org/#1}{\hspace*{0.1em}\raisebox{-0.45ex}{\includegraphics[width=1em]{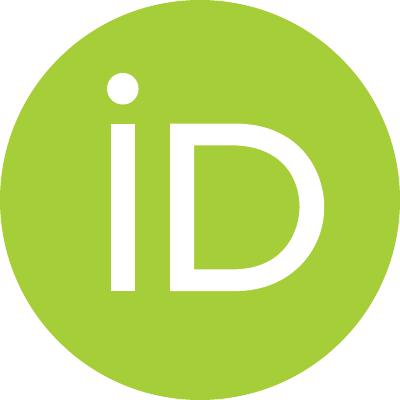}}}}

\def\sigmapmumu{\ensuremath{\PSigma^+ \to \Pp \mu^+ \mu^-}\xspace}

\def\pmumu{\ensuremath{\Pp \mu^+ \mu^-}\xspace}
\def\mpmumu{\ensuremath{m_{\Pp \mu^+ \mu^-}}\xspace}
\def\sigmapgamma{\ensuremath{\PSigma^+ \to \Pp \gamma}\xspace}
\def\sigmapxmumu{\ensuremath{\PSigma^+ \to \Pp X^0 (\to \mu^+ \mu^-)}\xspace}
\def\sigmapmumulfv{\ensuremath{\PSigma^+ \to \antiproton \mu^+ \mu^+}\xspace}
\def\sigmappiz{\ensuremath{\PSigma^+ \to \Pp \pi^0}\xspace}

\def\sigmappizero{\sigmappiz}

\def\kpipipi{\ensuremath{K^+ \to \pi^+ \pi^- \pi^+}\xspace}
\def\kpimumu{\ensuremath{K^+ \to \pi^+ \mu^- \mu^+}\xspace}

\def\mmumu{\ensuremath{m_{\mu^+\mu^-}}\xspace}

\def\mppi{\ensuremath{m_{p\pi^-}}\xspace}

\def\bujpsik{\ensuremath{B^+ \to J/\psi K^+}\xspace}
\def\mcorr{\ensuremath{m_{\rm Corr}}\xspace}

\hypersetup{
  colorlinks   = true, 
  urlcolor     = blue, 
  linkcolor    = blue, 
  citecolor    = red   
}

\ifEnableSectionTOCLinks
    \usepackage[explicit]{titlesec} 
    
    \let\oldcontentsline\contentsline
    \renewcommand

    \titleformat{\section}{\normalfont\Large\bf}{\hyperlink{tocsection.\thesection}{{\thesection} \parbox[t]{\dimexpr\textwidth-1pc}{#1}}}{1pc}{}

    \titleformat{\subsection}{\normalfont\bf}{\hyperlink{tocsubsection.\thesubsection}{{\thesubsection} \parbox[t]{\dimexpr\textwidth-1pc}{#1}}}{1pc}{}

    \titleformat{name=\section,numberless}[display]{}{}{0pt}{\normalfont\Huge\bfseries #1}
\fi

\usepackage{cite} 
\usepackage{mciteplus}

\def\NSigmappizeroRun1{\ensuremath{(1171 \pm 9)\times 10^{3}}\xspace}
\def\NTOTSigmaRun1{\ensuremath{ 10^{14}}\xspace}

\def\alphafullRun1{\ensuremath{(2.2 \pm 1.2)\times 10^{-9}}\xspace}
\def\expeventsRun1{\ensuremath{23 \pm 20}\xspace}
\def\signdefaultRun1{\ensuremath{4.1}\xspace}
\def\nsigmadefaultRun1{\ensuremath{ 10.2\,^{+\,3.9}_{-\,3.5}}\xspace}

\def\upperlimithcpninetyRun1{\ensuremath{1.4 \times 10^{-8}}\xspace}
\def\upperlimithcpninetyfiveRun1{\ensuremath{1.7 \times 10^{-8}}\xspace}

\def\candidateshcpRun1{\ensuremath{3.0\,^{+\,1.7}_{-\,1.4}}\xspace}
\def\fractionhcpRun1{\ensuremath{30\%}\xspace}

\def\brmeasuredRun1{\ensuremath{(2.2\,^{+\,1.8}_{-\,1.3})\times 10^{-8}}\xspace}
\def\brmeasuredsystRun1{\ensuremath{(2.2\,^{+\,0.9}_{-\,0.8}\,^{+\,1.5}_{-\,1.1})\times 10^{-8}}\xspace}

\def\NsignalRun2{\ensuremath{237 \pm 16}\xspace}

\def\NsignalTOSRuntwo{\ensuremath{96\pm 10}\xspace}
\def\NsignalTISRuntwo{\ensuremath{154\pm 13}\xspace}
\def\NnormTOSRuntwo{\ensuremath{(6.13 \pm 0.10) \times 10^3}\xspace}
\def\NnormTISRuntwo{\ensuremath{(4.746 \pm 0.031) \times 10^4}\xspace}
\def\brmeasuredRuntwo{\ensuremath{(1.08 \pm 0.17)\times 10^{-8}}\xspace}

\def\brmeasuredCombined{\ensuremath{(1.09 \pm 0.17)\times 10^{-8}}\xspace}

\def\alphatosruntwo{\ensuremath{(1.65^{\,+\,0.09}_{\,-\,0.16}{}^{\,+\,0.41}_{\,-\,0.41})\times 10^{-10}}\xspace}
\def\alphatisruntwo{\ensuremath{(6.81^{\,+\,0.29}_{\,-\,0.25}{}^{\,+\,0.85}_{\,-\,0.85})\times 10^{-11}}\xspace}
\def\brmeasuredTOSRuntwo{\ensuremath{(1.59^{\,+\,0.19}_{\,-\,0.23}{}^{\,+\,0.40}_{\,-\,0.40})\times 10^{-8}}\xspace}
\def\brmeasuredTISRuntwo{\ensuremath{(1.05^{\,+\,0.10}_{\,-\,0.10}{}^{\,+\,0.13}_{\,-\,0.13})\times 10^{-8}}\xspace}

\usepackage{longtable} 

\begin{document}

\renewcommand{\thefootnote}{\fnsymbol{footnote}}
\setcounter{footnote}{1}


\begin{titlepage}
\pagenumbering{roman}

\vspace*{-1.5cm}
\centerline{\large EUROPEAN ORGANIZATION FOR NUCLEAR RESEARCH (CERN)}
\vspace*{1.5cm}
\noindent
\begin{tabular*}{\linewidth}{lc@{\extracolsep{\fill}}r@{\extracolsep{0pt}}}
\ifthenelse{\boolean{pdflatex}}
{\vspace*{-1.5cm}\mbox{\!\!\!\includegraphics[width=.14\textwidth]{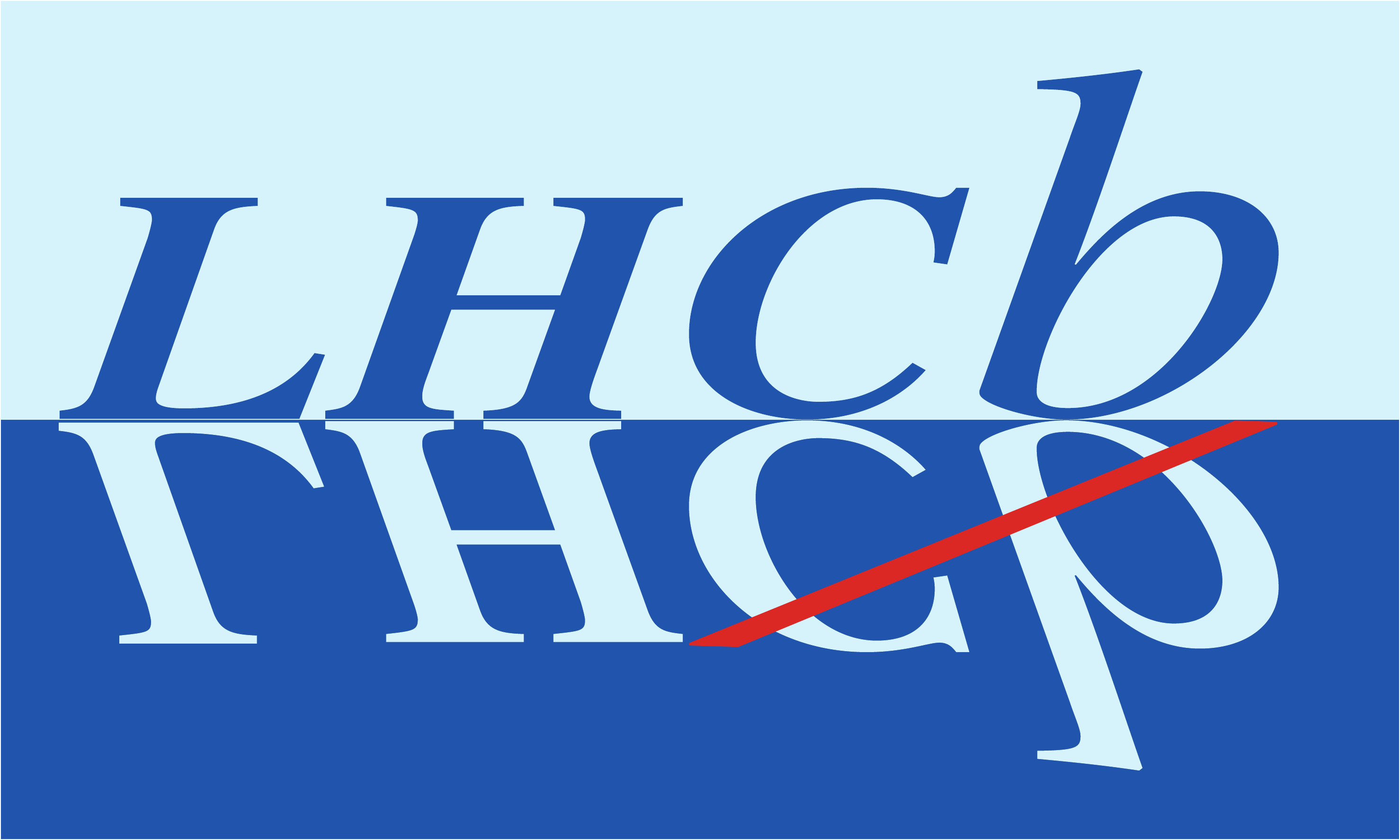}} & &}%
{\vspace*{-1.2cm}\mbox{\!\!\!\includegraphics[width=.12\textwidth]{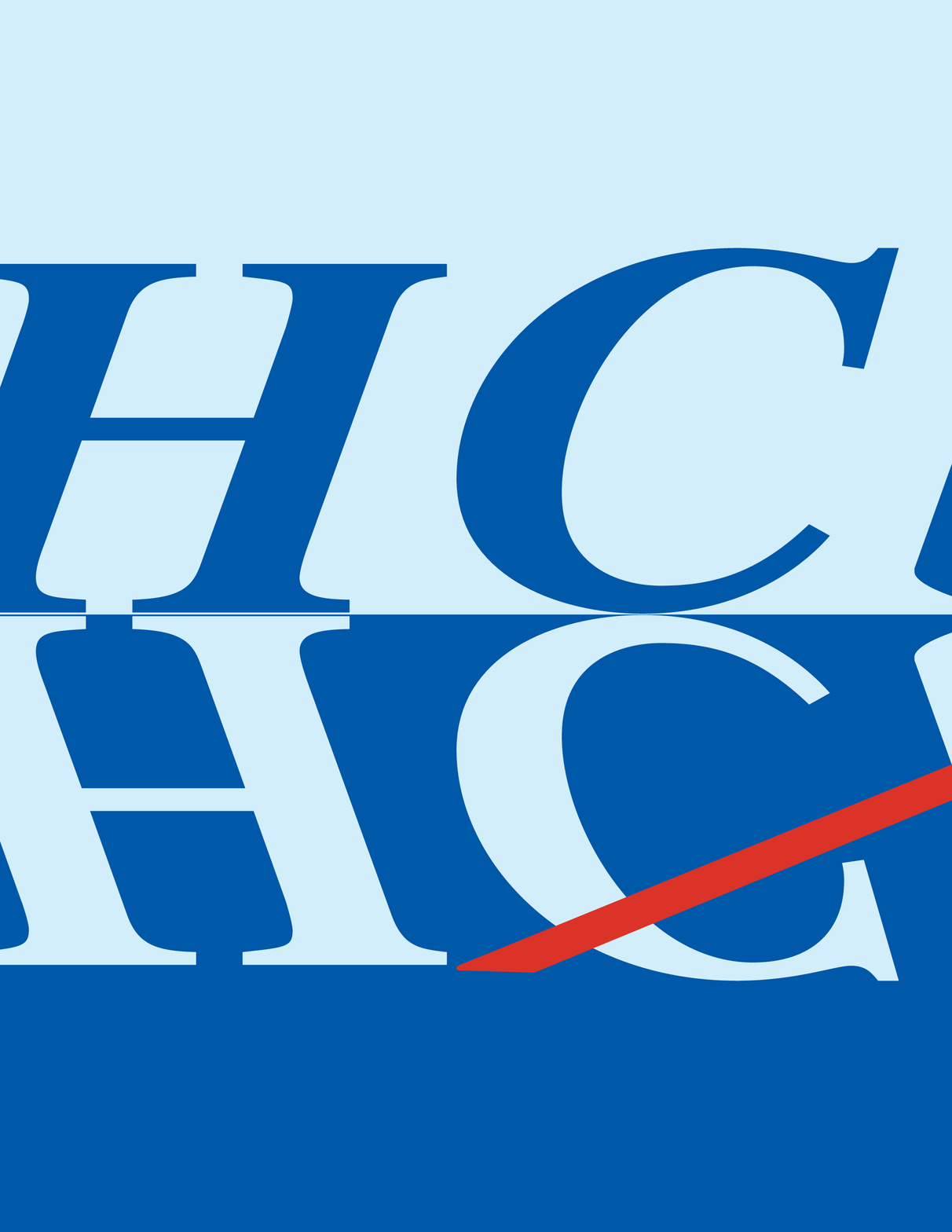}} & &}%
\\
 & & CERN-EP-2025-074\\  
 & & LHCb-PAPER-2025-002 \\  
 & & \today \\ 
 & & \\
\end{tabular*}

\vspace*{4.0cm}

{\normalfont\bfseries\boldmath\huge
\begin{center}
  \papertitle 
\end{center}
}

\vspace*{2.0cm}

\begin{center}
\paperauthors\footnote{Authors are listed at the end of this paper.}
\end{center}

\begin{abstract}
  \noindent
   The first observation of the $\Sigma^+ \to p \mu^+ \mu^-$ decay is reported with high significance using proton-proton collision data, corresponding to an integrated luminosity of $5.4\,\rm{fb}^{-1}$, collected with the LHCb detector at a centre-of-mass energy of 13~TeV.
 A yield of $237\pm 16$ $\Sigma^+ \to p \mu^+ \mu^-$ decays is obtained, where the uncertainty is statistical only.
 A branching fraction of $(1.08 \pm 0.17) \times 10^{-8}$ is measured, where the uncertainty includes statistical and systematic sources. No evidence of resonant structures is found in the dimuon invariant-mass distribution.  All results are compatible with Standard Model expectations.
 This represents the rarest decay of a baryon ever observed.
\end{abstract}

\vspace*{2.0cm}

\begin{center}
  Published in Phys. Rev. Lett. 135 (2025) 051801
\end{center}

\vspace{\fill}

{\footnotesize 
\centerline{\copyright~\papercopyright. \href{\paperlicenceurl}{\paperlicence}.}}
\vspace*{2mm}

\end{titlepage}


\newpage
\setcounter{page}{2}
\mbox{~}


\renewcommand{\thefootnote}{\arabic{footnote}}
\setcounter{footnote}{0}


\cleardoublepage


\pagestyle{plain} 
\setcounter{page}{1}
\pagenumbering{arabic}



The \sigmapmumu decay is a flavour-changing neutral-current (FCNC) process that is allowed only at loop level within the Standard Model (SM). 
The decay rate could be significantly modified by the presence of new physics~(NP) beyond the SM effects.
Short-distance contributions in the SM arise from box, $Z$- and electromagnetic-penguin processes, whose combined branching fraction is predicted to be $\mathcal{O}(10^{-12})$~\cite{He:2005yn}. While this is significantly smaller than the predicted long-distance contribution, it is worth noting that NP contributions are expected to manifest at short distance~\cite{He:2018yzu,Roy:2024hqg,Geng:2021fog}, potentially modifying the expected decay rate.
The long-distance SM contribution is calculated from weak nonleptonic decays $\decay{\Sigmap}{(N\pi)^+}$ and the subsequent reactions $\decay{(N\pi)^+}{p \gamma^{(*)}}$,\footnote{The inclusion of charge-conjugated processes is implied throughout.} where $N$ represents either a proton, $p$, or a neutron, $n$, and $\gamma^*$ is a virtual photon.
The branching fraction prediction contains an inherent eight-fold ambiguity due to the presence of four complex form factors studied in both relativistic and heavy-baryon chiral perturbation theory~($\chi$PT). 
A unitarity argument determines their imaginary parts~\cite{He:2005yn,He:2018yzu,Roy:2024hqg,Corrigan:1972sk,Bergstrom:1987wr,lyagin}, while the real components are predicted from the measured \sigmapgamma decay rate~\cite{BESIII:2023fhs}, which is responsible for a remaining four-fold degeneracy in each $\chi$PT approach.
The most recent theoretical predictions for the branching fraction $\mathcal{B}(\sigmapmumu)$ lie within the range $[1.2, 7.8]\times 10^{-8}$~\cite{Roy:2024hqg}.
Progress has also been made towards a lattice calculation of this branching fraction~\cite{Erben:2022tdu,Erben:2022igb}. 

Evidence for this channel was first found by the HyperCP experiment with a measured branching fraction of \mbox{$\mathcal{B}(\sigmapmumu) = (8.6^{\,+\,6.6}_{\,-\,5.4} \pm 5.5) \times 10^{-8}$}~\cite{Park:2005eka}, compatible with all SM predictions. 
Note that the measurement was based on three observed candidates with nearly identical dimuon invariant mass, close to the kinematic limit, suggesting an unexpected hint of structure with mass $m_{X^0}  = 214.3 \pm 0.5 \mevcc$. 
If confirmed, this would have pointed towards the decay of an intermediate particle into 
two muons, \ie a \mbox{\sigmapxmumu} decay. This result attracted significant theoretical attention
attempting to explain the origin of this hypothetical 
state~\cite{He:2006fr,He:2006uu,Gorbunov:2005nu,Demidov:2006pt,He:2005we,Geng:2005ra,Deshpande:2005mb,Chen:2007uv,Xiangdong:2007vv,Mangano:2007gi,Pospelov:2008zw}. 
In general, a pseudoscalar particle is favoured
over a scalar state,
with a lifetime in the order of $10^{-14} \sec$ estimated for the former.
Considerable experimental efforts have been made 
in order to search for this particle in other experiments and decay modes 
\cite{Love:2008aa,Tung:2008gd,Abazov:2009yi,Aubert:2009cp,Hyun:2010an,Abouzaid:2011mi,Ablikim:2011es,Lees:2014xha, LHCb-PAPER-2012-049, LHCb-PAPER-2015-036, LHCb-PAPER-2016-052, LHCb-PAPER-2017-038}.
The first search using \sigmapmumu decays since the HyperCP evidence was performed by LHCb with Run~1 proton-proton (\proton\proton) collision data~\cite{LHCb-PAPER-2017-049}, corresponding to an integrated luminosity of 3\invfb collected at a centre-of-mass energy $\sqs = 7\tev$. 
An excess of \nsigmadefaultRun1 events was observed with a significance of $\signdefaultRun1$ standard deviations~($\sigma$), corresponding to a branching fraction of
$\mathcal{B}(\sigmapmumu) =   \brmeasuredRun1 $, compatible with all SM predictions. 
The background-subtracted dimuon invariant-mass distribution was consistent with that of a phase-space (PHSP) simulation, leading to
an upper limit of \mbox{$\mathcal{B}(\Sigmap \to p X^0 ( \to \mu \mu)) <  \upperlimithcpninetyRun1$} at the 90\% confidence level for a hypothetical $X^0$ particle,  which disfavours the central value determined by the HyperCP collaboration.

This Letter presents the first observation of the \sigmapmumu decay. This analysis is performed with $pp$ collision data recorded by the LHCb experiment in 2016--2018 (Run~2) at $\sqrt{s} = 13\,\tev$, corresponding to an integrated luminosity of $5.4 \invfb$. A measurement of the \sigmapmumu branching fraction is reported, using the \sigmappizero decay as normalisation channel. The dimuon invariant-mass distribution for signal decays is also presented. 
This analysis follows a similar strategy to that performed using Run~1 data~\cite{LHCb-PAPER-2017-049}, with several improvements implemented, namely more efficient particle identification (PID), larger simulated samples and, most importantly, additional trigger selections that increase the signal efficiency by an order of magnitude, in addition to yield enhancements that naturally accompany increases in cross-section and luminosity. 
In order to avoid experimenter’s bias, the dimuon invariant-mass distribution was not examined and a random factor was kept in the branching fraction normalisation until the full analysis was finalised.

The LHCb detector is a single-arm forward spectrometer covering the pseudorapidity range 
$2 < \eta < 5$, 
described in detail in Refs.~\cite{LHCb-DP-2008-001,LHCb-DP-2014-002}.
Due to its lifetime of \mbox{$(8.018\pm 0.026)\times10^{-11}\sec$}~\cite{PDG2024}, the \Sigmap baryon can decay both inside the vertex detector, such that all decay products are reconstructed using the full tracking system (long tracks), or downstream of the vertex detector (downstream tracks). In this analysis, only long tracks are used. 
The short lifetime estimated for the $X^0$ particle would result in prompt signal production, hence, no attempt is made to distinguish the dimuon origin vertex from the decay vertex of the $\Sigmap$ baryon in this search.

The online event selection is performed by a trigger consisting of a hardware stage, using information from the calorimeter and muon systems, followed by two software stages, where a full event reconstruction is performed. Since 2016, two inclusive dimuon trigger selections have been added at the two software trigger stages, specifically designed to retain low transverse-momentum, \pt, track combinations whilst remaining within the strict processing time constraints imposed for the software trigger. In these selections, muon tracks are required to be inconsistent with originating from any primary vertex~(PV) and to have PID information consistent with their mass hypothesis. More details can be found in Refs.~\cite{LHCb-PUB-2017-023,LHCb:2018zdd}.
In addition, an exclusive trigger selection has been introduced for the \sigmapmumu decay channel. Candidate \Sigmap baryons are formed from combinations of a pair of oppositely charged muons and one proton candidate that are inconsistent with originating from any PV, with a good track-fit quality, and which form a good-quality vertex~\cite{LHCb-DP-2019-001}. The $\Sigmap$ candidate is required to have $\pt>500\,\mevc$, be consistent with originating from a PV and to have a significant flight distance.
The \sigmappizero decay, used as normalisation channel, is selected from a well-identified proton track and a $\pi^0$ candidate reconstructed in the two-photon final state from two clusters in the calorimeter.  For events with multiple candidates, all are accepted.

Given the large production rate of \Sigmap baryons in $pp$ collisions, the present search is conducted also including data selected at one or more trigger stages by other particles in the event. In the offline processing phase, trigger decisions are associated with reconstructed candidates. A trigger decision can thus be ascribed to the reconstructed candidate, the rest of the event or a combination of both. 
In the Triggered On Signal (TOS) sample the signal decay products are required to pass the muon or the hadron triggers, and the proton from \sigmappizero decays is required to pass the hadron trigger. In the Triggered Independently of Signal (TIS) sample, both channels are selected in events where their decay products are not necessary for the trigger decision. Candidates from the two samples are selected together, but the measurement of the signal branching fraction is performed separately for the TIS and TOS samples.
At the software trigger level, the signal selection already described is maintained for both samples. The normalisation channel is selected at the first software level by requiring a high \pt hadron, while a minimum-bias selection, with a fixed scale factor of $10^{-4}$ to limit its rate, is applied at the second software level.

Simulated samples are used to optimise the selection criteria, parametrise invariant-mass distributions, and characterise the detector resolution and efficiencies. These samples
are generated with the software described in Refs.~\cite{Sjostrand:2007gs,*Sjostrand:2006za,LHCb-PROC-2010-056,Lange:2001uf,davidson2015photos,Allison:2006ve, *Agostinelli:2002hh, LHCb-PROC-2011-006}.
The signal \sigmapmumu
decay is generated according to a PHSP model and weighted to reproduce the SM dimuon invariant-mass predicted spectrum~\cite{He:2018yzu,Roy:2024hqg}.
Simulated samples are weighted to reproduce the multiplicity distribution observed in data.

Two sources of background remain after the offline selection: the combinatorial background, composed from random associations of tracks present in the event; and the $\Lz$ background, consisting of genuine $\Lz \to p \pi^-$ decays where the pion is misidentified as a muon, combined with a third unrelated track identified as a muon. 
No other kinds of background contribute, owing mainly to the very small energy of $39.8\mev$ available in the \sigmapmumu reaction.
Misidentified decays from other hadrons also do not contribute.
Background from four- or more-body final states with unreconstructed particles could contribute, but would not peak in the \mpmumu distribution and thus is included in the combinatorial background. 
Finally, background including duplicate tracks from the same particle hits is rejected offline with a requirement on the minimum angle between each pair of tracks. Additional details on the background sources are reported in Appendix~\ref{app:background}.

To further reduce the background, a multivariate classifier is devised based on a Boosted Decision Tree (BDT) algorithm ~\cite{Breiman,AdaBoost} implemented in the TMVA toolkit~\cite{Hocker:2007ht,TMVA4}. 
This BDT algorithm combines different kinematic and geometric variables.
The BDT is trained using the signal simulated sample and the data sidebands as a proxy for combinatorial background considering candidates in the mass range $\mpmumu < 1173\mevcc$ or $1205<\mpmumu<1400\mevcc$. Under the $p \pi^-$ hypothesis, a veto on the $p\mu^-$ mass is applied in both samples around the known $\Lz$ mass value~\cite{PDG2024} ($\Lz$ veto) to enforce training against combinatorial background only.  
To avoid overtraining, the $k$-folding technique~\cite{Blum:1999:BHB:307400.307439}, with $k=9$, is applied. 
The BDT output ranges from zero, for background-like candidates, to unity, for signal-like candidates; the corresponding distributions for signal simulation and data are shown in the Supplemental Material~\cite{supplemental}. 
The data is divided into a sample with the $\Lz$ veto and its complementary sample, consisting mostly of the $\Lz$ background, where the BDT distribution is seen to be very similar for both background sources.
The same BDT classifier is applied to a sample of ``same-sign" \sigmapmumulfv candidates in data, where a signal would have to violate lepton-number conservation, verifying that fake structures in the background are not created in either the \mpmumu or \mmumu distributions.

The final selection is based on the BDT output, 
the muon and proton particle-identification variables~\cite{LHCb-DP-2018-001}, 
and the width of the \Lz veto window.
Criteria on these variables are optimised on a four-dimensional grid to give the largest significance, defined as ${N_{\rm S}}/{\sqrt{N_{\rm S}+N_{\rm B}}}$, where $N_{\rm S}$ is the expected signal and $N_{\rm B}$ the expected background yield.
The $N_{\rm S}$ estimate is based on a preliminary fit to data after tight selection criteria are applied along with the signal efficiency obtained from simulation. 
The $N_{\rm B}$ estimate is the sum of two contributions: one obtained from a fit to the \mpmumu sidebands for the combinatorial background; and another to estimate the residual \Lz  background based on a fit to the $\mppi$ distribution without the \Lz veto ($\pm 10 \mevcc$ around the known \Lz mass~\cite{PDG2024}).
The optimised requirements on the BDT output, particle identification and \Lz veto are applied equally to the TIS and TOS samples. 

\begin{figure}[tbp]
\centering
\includegraphics*[width=0.483\textwidth]{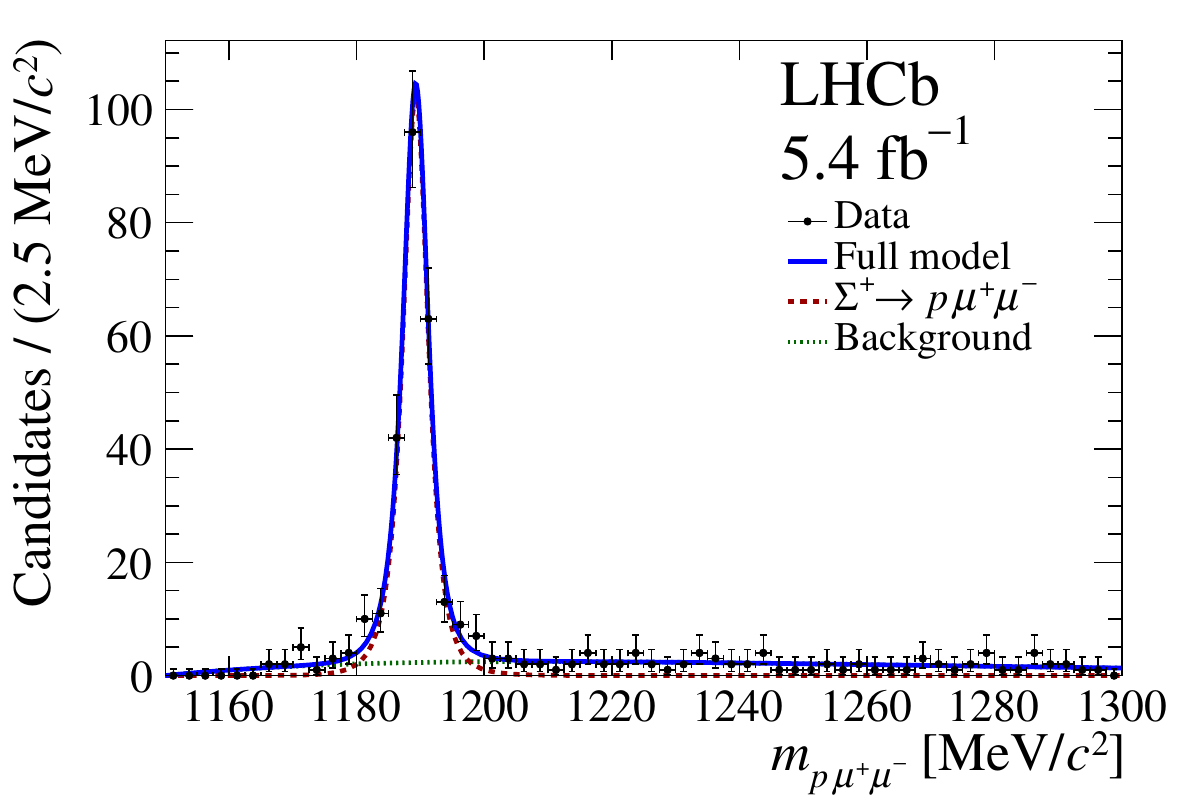}
\caption{ \label{fig:mpmumu} Distribution of the invariant mass of \sigmapmumu candidates in a restricted mass range with the result of the extended unbinned maximum-likelihood fit also shown (blue solid line). The signal (red dashed line) and background (green dotted line) components are also illustrated.
    The full range is shown in the Supplemental Material~\cite{supplemental}.}
\end{figure}

The \mpmumu distribution for candidates satisfying the final selection criteria is shown in Fig.~\ref{fig:mpmumu}, in which 
a clear peak at the $\Sigmap$ mass is observed with a small residual background.
An extended unbinned maximum-likelihood fit~\cite{Barlow:1990vc} is performed to the selected candidates.  
The signal component is described by a Hypatia function~\cite{MartinezSantos:2013ltf}, with the $z$ parameter fixed to zero. The remaining parameters are obtained from a fit to the simulated sample and fixed in the fit to data, while the peak position and resolution parameters are left free to vary.
The background is described by 
a modified Argus function~\cite{ARGUS:1990hfq}, where the threshold parameter is fixed to the sum of the final-state masses and the remaining parameters are free to vary. 
The data and the result of the fit are shown in Fig.~\ref{fig:mpmumu}; a signal yield of 
$N_{\sigmapmumu} = \NsignalRun2$ is obtained, 
where the uncertainty is statistical only. 
This result constitutes the first observation of the \sigmapmumu decay, obtained with overwhelming significance. 
The fit to data is repeated, releasing all parameters of the signal function that are fixed from simulation in the baseline model. 
The variation of the signal yield is negligible, hence no systematic uncertainty is assigned. 
A similar amount of $\Sigmap$ and $\Sigmabarm$ decays are seen in the sample. 

\begin{figure}[tbp]
    \centering
    \includegraphics[width= 0.48\textwidth]{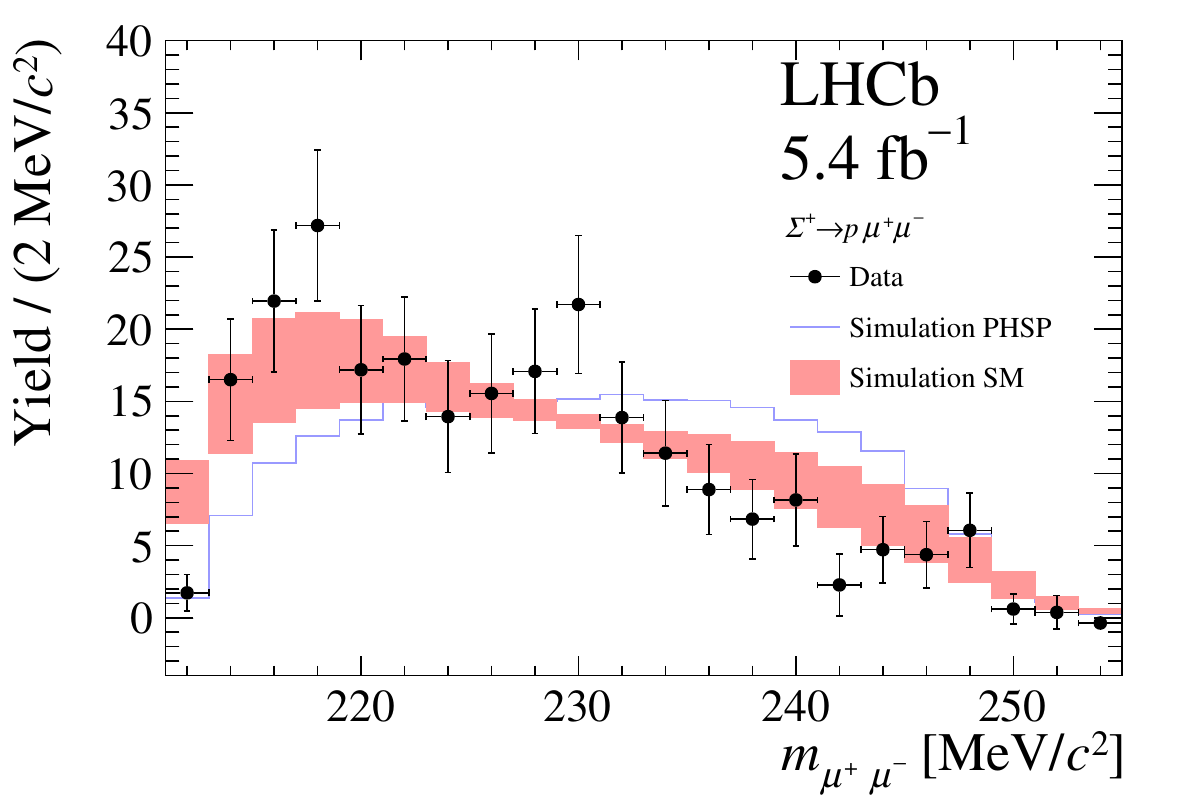}
    \caption{Distribution of background-subtracted dimuon invariant mass for \sigmapmumu candidates in data compared with simulation. 
LHCb PHSP simulation is shown as is (blue line), and weighted according to the SM amplitude~\cite{He:2018yzu,Roy:2024hqg} (red band). 
Note: the distributions in this figure are not corrected for the efficiencies. }
    \label{fig:dimuon}
\end{figure}

The distribution of the dimuon invariant mass is shown for data in Fig.~\ref{fig:dimuon} after background subtraction. 
The background is subtracted using per-event signal weights derived with the \textit{sPlot} method~\cite{Pivk:2004ty} using \mpmumu as the discriminant variable. The \mpmumu and \mmumu variables are found to be uncorrelated aside from the higher border of the kinematics space.  A consistent distribution is obtained when performing the unbinned maximum-likelihood fit described earlier in intervals of the dimuon invariant mass. No significant peaking structures are visible in the data distribution. The data is compared to the distribution in simulation with the PHSP model and with the SM weighted distribution, with theoretical uncertainty due to the mentioned parametric ambiguity shown as a red band in Fig.~\ref{fig:dimuon}. When comparing the data to this SM simulation, rather good agreement is achieved in the full range, with a qualitative preference for the lowest predicted degenerate branching fraction~\cite{Roy:2024hqg}.

A scan for possible resonant structure in the dimuon invariant mass is performed by selecting candidates in the \pmumu invariant mass within twice the signal resolution of the known \Sigmap mass, using the same method as in the Run~1 analysis~\cite{LHCb-PAPER-2017-049}, and detailed in Appendix~\ref{app:scan}.
Steps of half the resolution on the dimuon invariant mass, $\sigma(m_{\mu^+\mu^-})$, are considered in this scan, following the method outlined in Ref.~\cite{Williams:2015xfa}.  The value of $\sigma(m_{\mu^+\mu^-})$ varies in the range \mbox{[0.5, 2.0]\mevcc} depending on the dimuon invariant mass.
For each step, the putative signal is estimated in a window of $\pm 1.5\times\sigma(m_{\mu^+\mu^-})$ around the considered particle mass,
 while the background is estimated from the lower and upper mass sidebands contained in the range $[1.5,4.0]\times\sigma(m_{\mu^+\mu^-})$
for the same mass. Only one of the two sidebands is considered when the other is outside the allowed kinematic range. 
The local $p$-value of the background-only hypothesis is shown in Appendix~\ref{app:scan} as a function of the dimuon invariant mass. No significant signal is found; the most significant point occurs at 247.06 \mevcc, with a $p$-value of 3.5\%.
When considering a putative candidate with a mass $m_{X^0} = 214.3 \mevcc$~\cite{Park:2005eka}, the fractional contribution to all candidates in the \pmumu signal mass window is 5\% and the difference with respect to the expected background from the \mmumu sidebands (\ie nonresonant) is $-1$ candidates.

The \sigmapmumu decay is normalised to  the \sigmappizero  decay in order to measure its branching fraction as
\begin{equation}
\label{eq:br_norm}
\mathcal{B}(\sigmapmumu) = \frac{\varepsilon_{\rm{Norm}}}{\varepsilon_{\rm{Sig}}} 
\frac{N_{\rm{Sig}}}{N_{\rm{Norm}}} \mathcal{B}(\sigmappizero) = \alpha  N_{\rm{Sig}}\,,
\end{equation}
where $\varepsilon$ and $N$ are the efficiency and yield of the indicated channel,  \mbox{$\mathcal{B}(\sigmappizero) = (51.57 \pm 0.30)\%$~\cite{PDG2024}}, and $\alpha$ is the single event sensitivity.
The efficiencies in Eq.~\eqref{eq:br_norm} are factorised into different categories for ease of estimation and each is evaluated with respect to the previous steps: detector acceptance, reconstruction and selection, PID, and trigger, and the yields are estimated from fits to their respective innvariant-mass distributions as described later. 

The acceptance, reconstruction, and selection efficiencies are obtained from simulation.
Possible residual differences between data and simulation in the tracking efficiencies are determined using control data samples~\cite{LHCb-DP-2013-002}. The uncertainty on the theoretical spectrum is propagated to the efficiency as a systematic uncertainty amounting to 2.5\%. 
The PID efficiencies are determined from data using samples of kinematically identified charged particles from \bujpsik decays for the muons, and $\Lz \to \Pp \Ppi^-$ decays for the protons~\cite{LHCb-DP-2018-001}.
 Systematic uncertainties of about 1\% and 5\% are associated with the PID calibration of muons and protons, respectively.
Tracking efficiencies are determined in bins of 
$p$  and $\eta$ and compared to those from simulation. The weighted average correction is calculated and applied to signal and normalisation simulated samples. A total systematic uncertainty of about 5\% is associated with this calibration.
The $\pi^0$ reconstruction efficiency is calibrated using $B^+ \to J/\psi K^{\ast +} (\to K^+ \pi^0)$ and $B^+\to J/\psi K^+$ decays~\cite{LHCB-PAPER-2012-022, LHCb-PAPER-2012-053,Belyaev:2015nlt,Govorkova:2015vqa}, for which a systematic uncertainty of 7\%, mostly due to the branching fractions of the calibration channels,  is assigned.

The hardware hadron trigger efficiencies, for both signal and normalisation channels, are evaluated with two methods: the \textsc{Tistos} method~\cite{LHCb-DP-2012-004,Tolk:1701134} in data, 
using the normalisation channel, and independent estimates based on control samples in data, obtained with similar methods as in Ref.~\cite{LHCb-PUB-2011-026}; from the comparison of the two methods a 12\%  relative systematic uncertainty is assigned.
The hardware muon trigger efficiencies for the signal are evaluated similarly to Refs.~\cite{LHCb-PAPER-2019-038,LHCb-PAPER-2022-035}. There, given the lack of other low-\pt calibration channels, the $\KS\to \pi \mu \nu$ channel was used, with the pion decaying in flight into a muon. 
In this analysis, the \kpipipi channel is used, where two opposite-sign pions are identified as muons. 
Following Refs.~\cite{LHCb-PAPER-2019-038,LHCb-PAPER-2022-035}, a systematic uncertainty is assigned to the TOS sample
which includes data-simulation agreement and the method verification, and amounts to 22\%. 
The efficiency of the TIS hardware trigger category is estimated with \sigmappizero decays in data with the \textsc{Tistos} method and corrected for the residual correlation with the decay kinematics. 
A systematic uncertainty of 6\% is assigned. 

The efficiency of the requirement on the BDT output is calibrated directly on \mbox{\sigmapmumu} decays by comparing the data and simulation distribution above the chosen value. To account for the small disagreement between data and simulation, the data distribution is extrapolated to lower BDT output values to obtain an efficiency correction for the simulation. The distribution is well described by a first-order polynomial, however a second-order is also employed with the difference assigned as a systematic uncertainty. 

\begin{figure}
    \centering
    \includegraphics[width=0.48\textwidth]{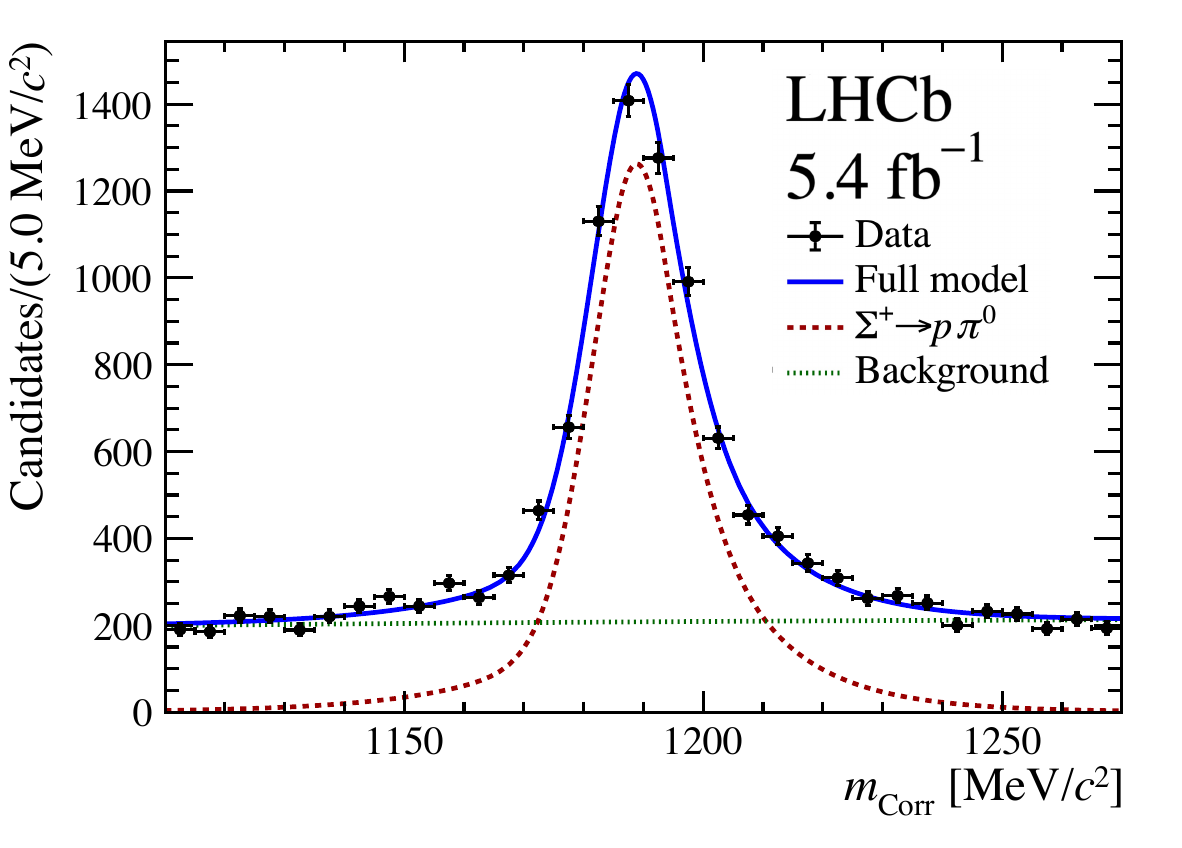}
    \caption{Distribution of the corrected mass for \sigmappizero decays in data for the TOS sample with the result of an extended unbinned maximum-likelihood fit  (blue solid line). The \sigmappizero  (red dashed line) and background (green dotted line) components are also shown.  }
    \label{fig:normfit}
\end{figure}

The yield of the \sigmappizero channel is obtained through an unbinned maximum-likelihood fit 
to the distribution of the corrected mass, defined as 
$m_{\rm Corr} \equiv m_{p\gamma \gamma} - m_{\gamma\gamma} + m_{\pi^0}^{\rm PDG}$. Here, $m_{p\gamma\gamma}$ is the invariant mass of the proton and two photons, $m_{\gamma\gamma}$ that of the  two photons from which the $\pi^0$ is reconstructed,  and $m_{\pi^0}^{\rm PDG}$ is the known $\pi^0$ mass~\cite{PDG2024}. 
The \sigmappizero probability distribution function is composed of   the sum of a Gaussian 
and a Crystal Ball function~\cite{Skwarnicki:1986xj},  with power-law tails on both sides. 
The background component is described by a Chebyshev second-degree polynomial of the first kind. The \sigmappizero tail parameters are obtained from simulation, while the remaining parameters are free in the fit. 
The distribution of \mcorr is shown in Fig.~\ref{fig:normfit}; a total of \NnormTOSRuntwo and \NnormTISRuntwo \sigmappizero candidates is obtained for the TOS and TIS samples, respectively.
Studies of the \sigmappizero yield using alternative background parametrisations show variations that are negligible with respect to the statistical uncertainty, hence no additional systematic uncertainty is associated to this fit.
With Eq.~\eqref{eq:br_norm}, the single-event sensitivities of the two samples are $\alpha_{\rm TOS} = \alphatosruntwo$ and $\alpha_{\rm{TIS}}= \alphatisruntwo$, respectively, where the first uncertainty is statistical and the second systematic. 

The fit to the \sigmapmumu invariant-mass distribution is repeated separately for the TOS and TIS samples, with total yields of 
$\NsignalTOSRuntwo$ and $\NsignalTISRuntwo$, respectively.
These correspond to the measured branching fractions 
$\mathcal{B}_{\rm TOS} = \brmeasuredTOSRuntwo$ and $\mathcal{B}_{\rm TIS} = \brmeasuredTISRuntwo$, respectively, 
where the first uncertainty is statistical and the second systematic. 
The two branching fractions are then combined as a weighted average, factorising common efficiencies to take correlations into account. 
The resulting branching fraction is 
\[
\mathcal{B}(\sigmapmumu) = \brmeasuredRuntwo,
\]
where the uncertainty includes statistical and systematic sources.
This result is in agreement with the SM predictions~\cite{He:2005yn,He:2018yzu,Roy:2024hqg}.
The combination with the value obtained in Run~1~\cite{LHCb-PAPER-2017-049}, which has a large uncertainty, results in $\mathcal{B}(\sigmapmumu) = \brmeasuredCombined$.

In summary, the \sigmapmumu decay is observed with very high significance in data collected in Run~2 by the LHCb experiment in $pp$ collisions, with a yield of \mbox{$N_{\sigmapmumu} = \NsignalRun2$}.
No structure is seen in the dimuon invariant-mass distribution, which is compatible with expectations from the SM. 
The \sigmapmumu branching fraction, when combined with the Run~1 result, is measured to be 
\[
\mathcal{B}(\sigmapmumu) = \brmeasuredCombined\\,
\]
which is compatible with the SM and represents
the rarest baryon decay ever observed. From the combined experimental and theoretical $z$-scores, this result favours the lowest predicted degenerate branching fraction~\cite{Roy:2024hqg}, which is also qualitatively favoured by the structure measured in the $m_{\mu^+\mu^-}$ distribution.
Comparing the measured branching fraction with the eight theoretical predictions~\cite{ Roy:2024hqg}, the next largest is disfavoured by at least $3.7\sigma$, while the remaining two are rejected at over $6.1\sigma$. This comparison also shows that, in general, heavy-baryon $\chi$PT predictions are disfavoured by at least $3.1\sigma$ compared to relativistic $\chi$PT predictions.

With the collected signal yield, a measurement of additional observables such as the differential branching fraction, charge-parity symmetry violation and forward-backward asymmetries is envisaged and left for a future publication.

\section*{Acknowledgements}
%
%
\noindent We express our gratitude to our colleagues in the CERN
accelerator departments for the excellent performance of the LHC. We
thank the technical and administrative staff at the LHCb
institutes.
We acknowledge support from CERN and from the national agencies:
ARC (Australia);
CAPES, CNPq, FAPERJ and FINEP (Brazil); 
MOST and NSFC (China); 
CNRS/IN2P3 (France); 
BMBF, DFG and MPG (Germany); 
INFN (Italy); 
NWO (Netherlands); 
MNiSW and NCN (Poland); 
MCID/IFA (Romania); 
MICIU and AEI (Spain);
SNSF and SER (Switzerland); 
NASU (Ukraine); 
STFC (United Kingdom); 
DOE NP and NSF (USA).
We acknowledge the computing resources that are provided by ARDC (Australia), 
CBPF (Brazil),
CERN, 
IHEP and LZU (China),
IN2P3 (France), 
KIT and DESY (Germany), 
INFN (Italy), 
SURF (Netherlands),
Polish WLCG (Poland),
IFIN-HH (Romania), 
PIC (Spain), CSCS (Switzerland), 
and GridPP (United Kingdom).
We are indebted to the communities behind the multiple open-source
software packages on which we depend.
Individual groups or members have received support from
Key Research Program of Frontier Sciences of CAS, CAS PIFI, CAS CCEPP, 
Fundamental Research Funds for the Central Universities,  and Sci. \& Tech. Program of Guangzhou (China);
Minciencias (Colombia);
EPLANET, Marie Sk\l{}odowska-Curie Actions, ERC and NextGenerationEU (European Union);
A*MIDEX, ANR, IPhU and Labex P2IO, and R\'{e}gion Auvergne-Rh\^{o}ne-Alpes (France);
Alexander-von-Humboldt Foundation (Germany);
ICSC (Italy); 
Severo Ochoa and Mar\'ia de Maeztu Units of Excellence, GVA, XuntaGal, GENCAT, InTalent-Inditex and Prog. ~Atracci\'on Talento CM (Spain);
SRC (Sweden);
the Leverhulme Trust, the Royal Society and UKRI (United Kingdom).

\clearpage

\setcounter{equation}{0}

\clearpage
\appendix 
\section{End matter}

\subsection{Additional information on background sources}
\label{app:background}

As described in the main text, no other background source is expected in the selected sample aside from those already included: combinatorial and from misidentified $\Lz \to p \pi^-$ decays combined with an additional track.
The reason only these background sources can contribute is due to the extremely small phase space of the decay. 
The difference between the $\Sigma^+$ mass and the sum of the final-state particles masses is $39.8 \mevcc$~\cite{PDG2024}.
This implies that very few decays can mimic this final state at the same mass. 
In particular, the $\PSigma^+ \to p \pi^+\pi^-$ or $\PSigma^+ \to \Lz \mu^+ \nu_{\mu}$ decays are forbidden by energy conservation.
Given the low energy in the \sigmapmumu reaction, no background is expected from meson decays. 
For example, the \mbox{\kpipipi} and \kpimumu decay peaks in the invariant mass of the final state particles, \mpmumu, is shifted considerably higher in mass. Higher-mass mesons would be shifted even more.
As far as background from baryon decays is concerned, where the final-state proton is correctly identified, the $\Sigmap$ triplet is the lightest state with a significant lifetime, hence all other possible background sources will have a mass that is considerably larger.

Regarding possible residual \Lz background, no significant contribution is expected after the full selection including the \Lz veto.
The distribution of the $p\mu^-$ invariant mass in the $p\pi^-$ hypothesis is shown in the Supplemental Material~\cite{supplemental} for \sigmapmumu candidates in data 
    within a $\pm 6 \mevcc$ window from the $\Sigmap$ mass peak, without the \Lz veto. 
    The residual background coming from \Lz decays 
    outside of the veto represents only a small tail distributed along the \mpmumu mass, taken into account by the combinatorial background.

\subsection{Fit to the \sigmapmumu invariant mass}
\label{app:fit}

\begin{figure}[!tbh]
\centering
\includegraphics*[width=0.483\textwidth]{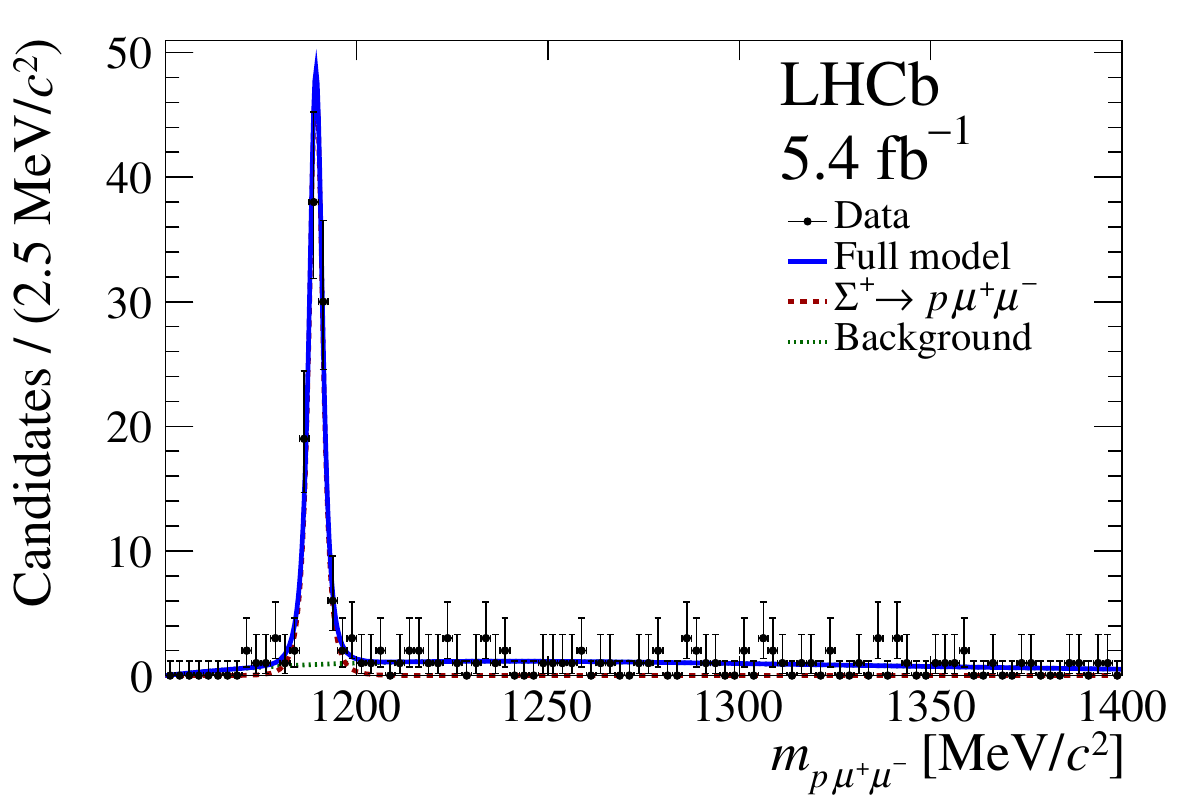}
\includegraphics*[width=0.483\textwidth]{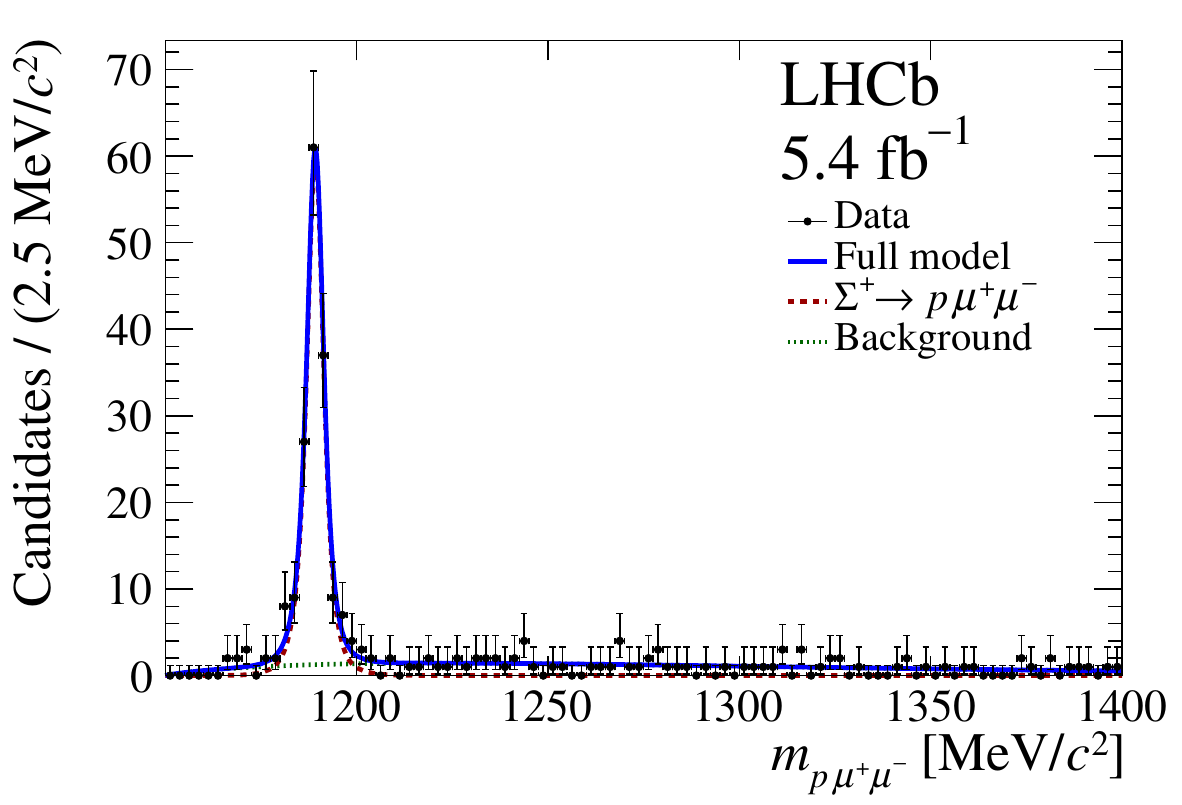}

\caption{ \label{fig:mpmumu_tistos} Distribution of the invariant mass of (left) \sigmapmumu TOS candidates with the result of an extended unbinned maximum-likelihood fit also shown and (right) corresponding figure for the TIS candidates.}
\end{figure}

The unbinned maximum-likelihood fit to the \pmumu invariant mass on the data sample divided into TOS and TIS samples is reported in Fig.~\ref{fig:mpmumu_tistos}.

\subsection{Scan for a structure in the dimuon invariant mass }
\label{app:scan}

\begin{figure}[!tbp]
\includegraphics[width = 0.49\textwidth]{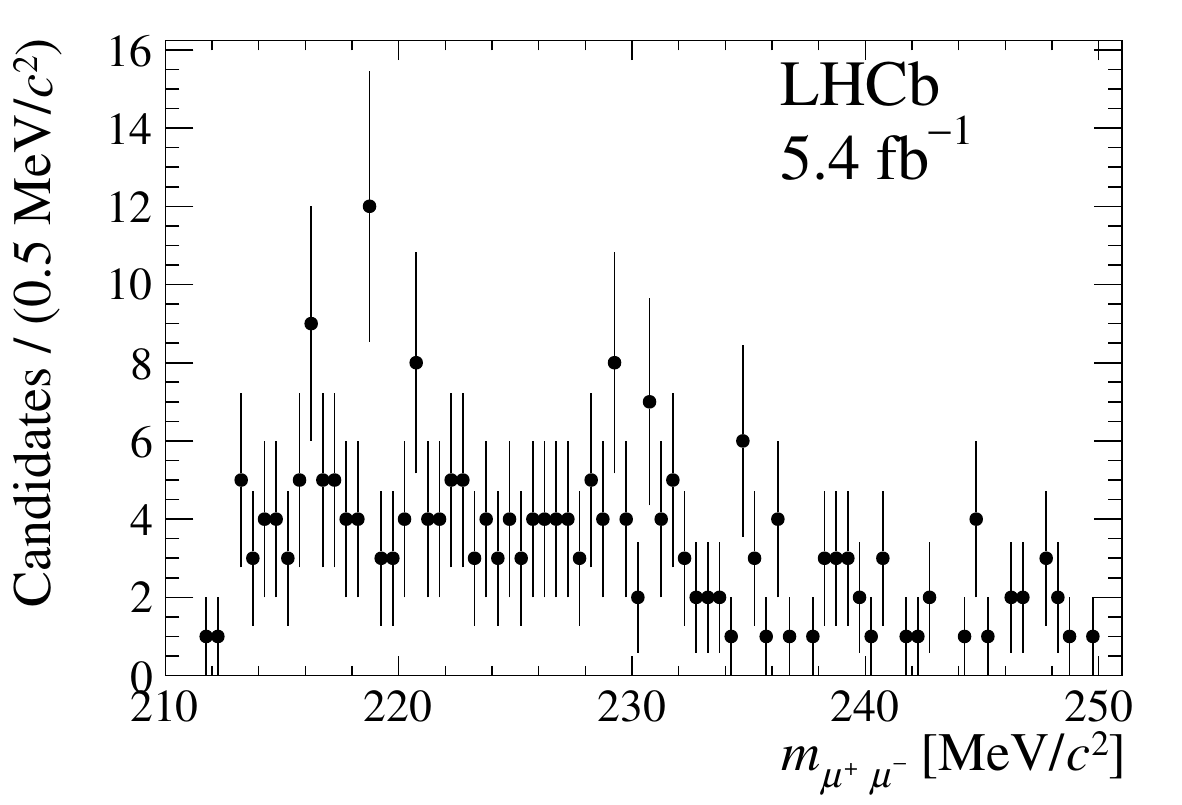}
 \includegraphics[width = 0.49\textwidth]{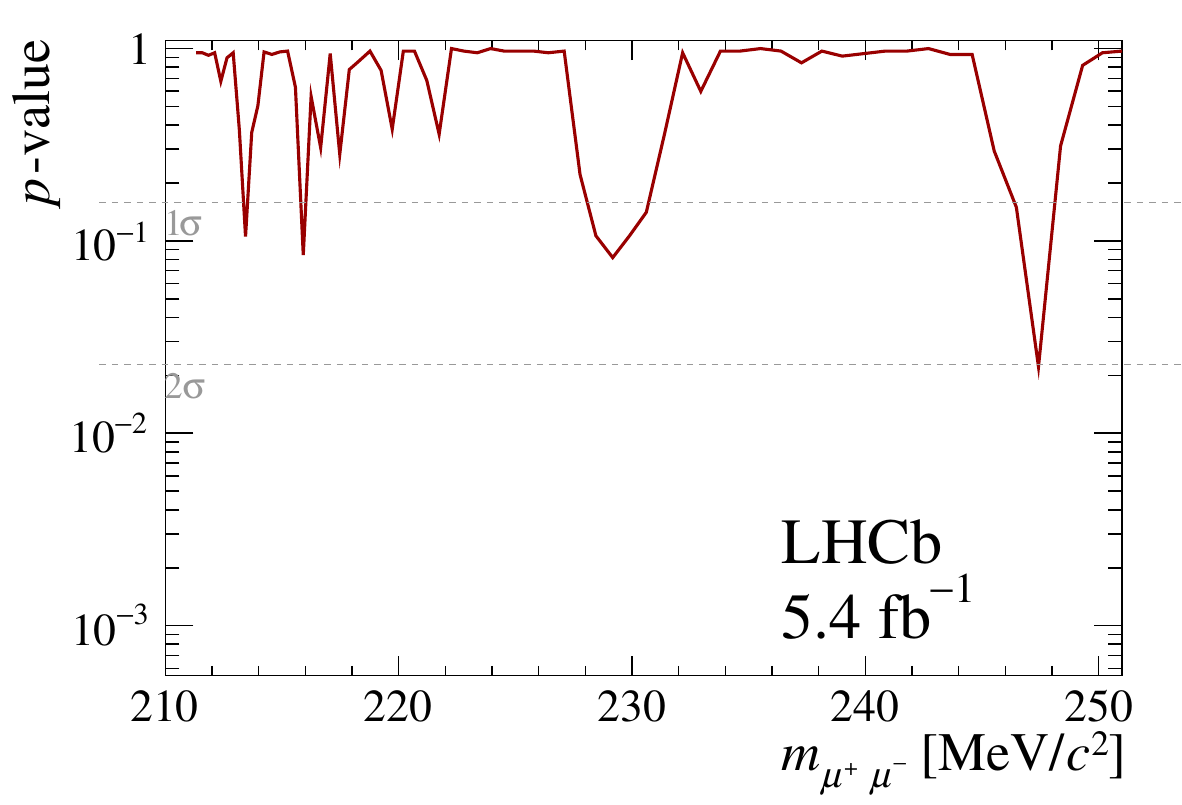}
 \caption{Distribution of (left) the dimuon invariant mass of \sigmapmumu decays in a signal region of $\pm 2$ times the resolution on \mpmumu, for which no background subtraction is applied. The local $p$-value (right) is shown in each dimuon invariant-mass window, obtained as described in the text.
The horizontal dashed lines correspond to the $p$-values of one and two standard deviations.}
\label{fig:dimuonscan}
\end{figure}

A scan for a possible resonant structure in the dimuon invariant mass is performed as described in the main text. 
The distribution of \sigmapmumu candidates in the signal region as a function of the dimuon invariant mass is shown in Fig.~\ref{fig:dimuonscan}(left).
The mass resolution $\sigma(m_{\mu^+\mu^-})$ as a function of the dimuon invariant mass is shown in the Supplemental Material~\cite{supplemental}.
The local $p$-value of the background-only hypothesis obtained from the scan for possible structure is shown in Fig.~\ref{fig:dimuonscan}(right) as a function of the dimuon invariant mass.

\subsection{Fit to the \sigmappizero invariant mass}

\begin{figure}[tbph]
\centering
\includegraphics*[width=0.483\textwidth]{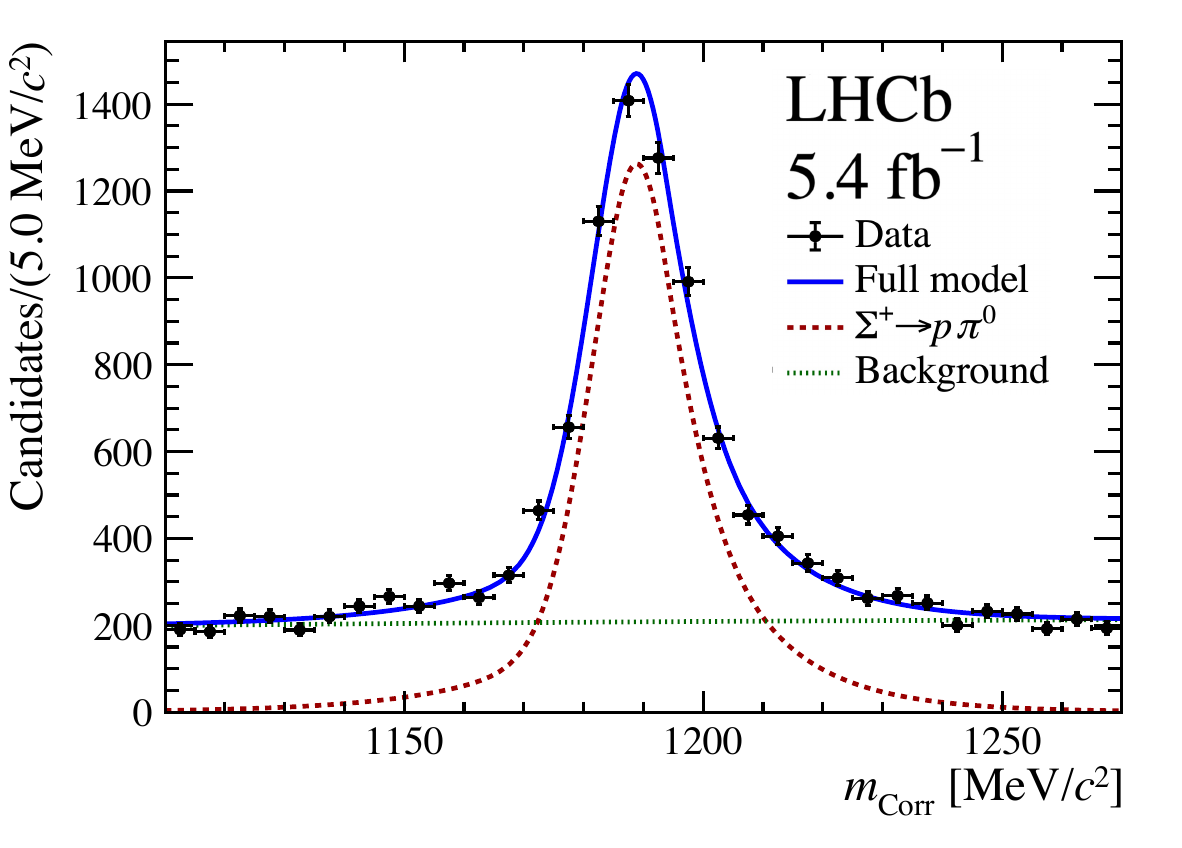}
\includegraphics*[width=0.483\textwidth]{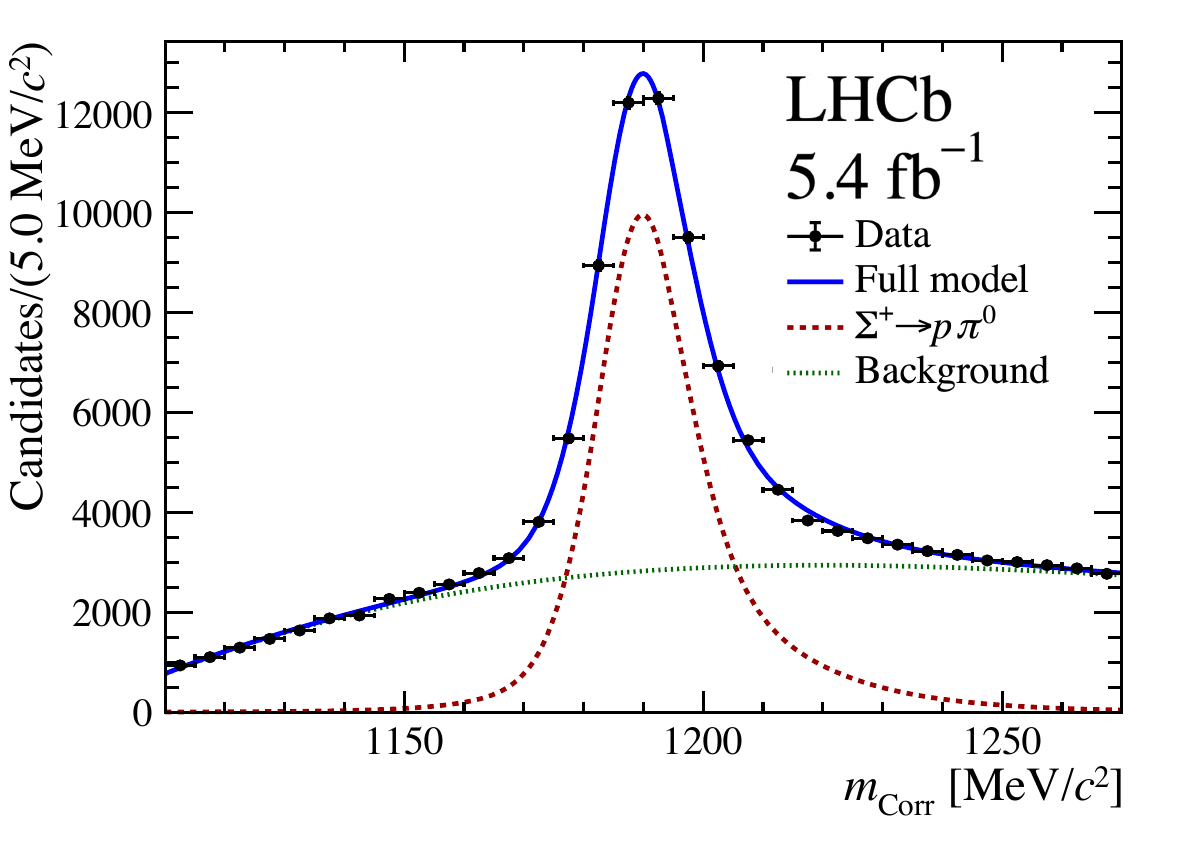}
\caption{ \label{fig:mppizero_tistos} Distribution of the corrected invariant mass of (left) $p\pi^{0}$ TOS candidates with the result of an extended maximum-likelihood fit also shown and (right) corresponding figure for the TIS candidates.}
\end{figure}

The extended maximum-likelihood fits to the corrected $p\pi^{0}$ invariant-mass distribution in data for the TIS and TOS samples are reported in Fig.~\ref{fig:mppizero_tistos}.

\clearpage

\clearpage
\section{Supplemental Material}
\label{sec:supplemental}

\begin{figure}[!h]
\centering
\includegraphics[width=0.48\textwidth]{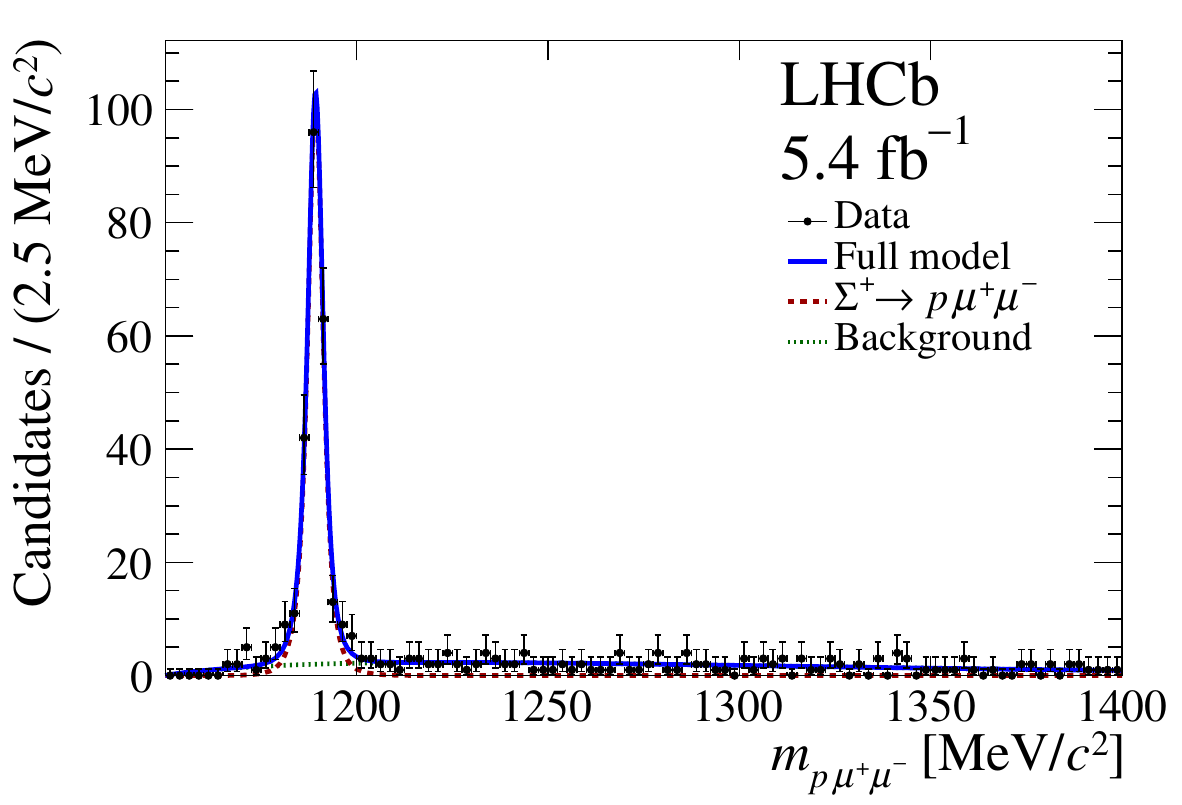}
\caption{ \label{fig:mpmumu_fullrange} Distribution of the invariant mass of \sigmapmumu candidates in the full fit range with the result of an extended unbinned maximum-likelihood fit also shown (blue solid line).}
\end{figure}

The unbinned maximum-likelihood fit to the \pmumu invariant mass, as described in the main text, is reported in its full fit range in Fig.~\ref{fig:mpmumu_fullrange}.

\begin{figure}[!tbp]
    \centering
    \includegraphics[width = 0.5 \textwidth]{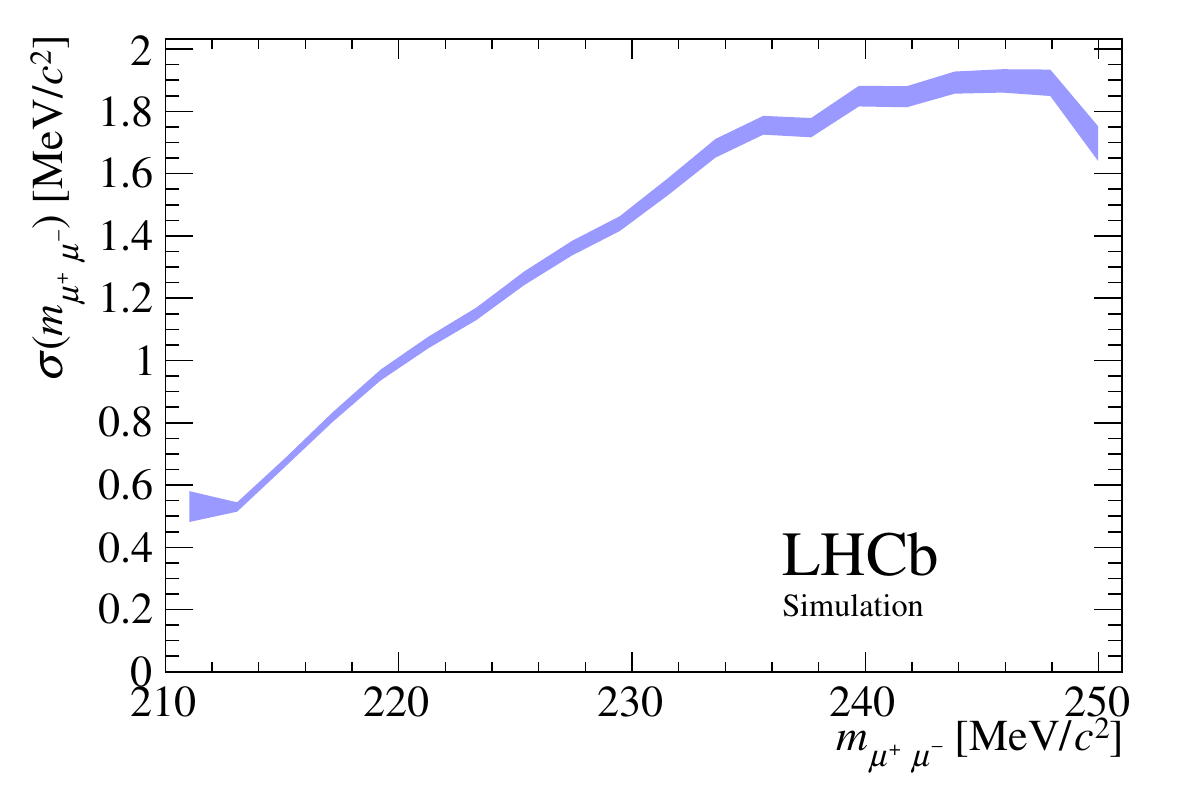}
    \caption{Dimuon invariant-mass resolution $\sigma(m_{\mu^+ \mu^-})$ for $\Sigma^+ \to p \mu^+ \mu^-$  candidates versus the dimuon mass.}
    \label{fig:dimuonres}
\end{figure}

Figure~\ref{fig:dimuonres} reports the resolution on the dimuon invariant mass, as obtained from simulation, as a function of the dimuon invariant mass in the full signal range. 

\begin{figure}[!bp]
    \centering
    \includegraphics[width = 0.5\textwidth]{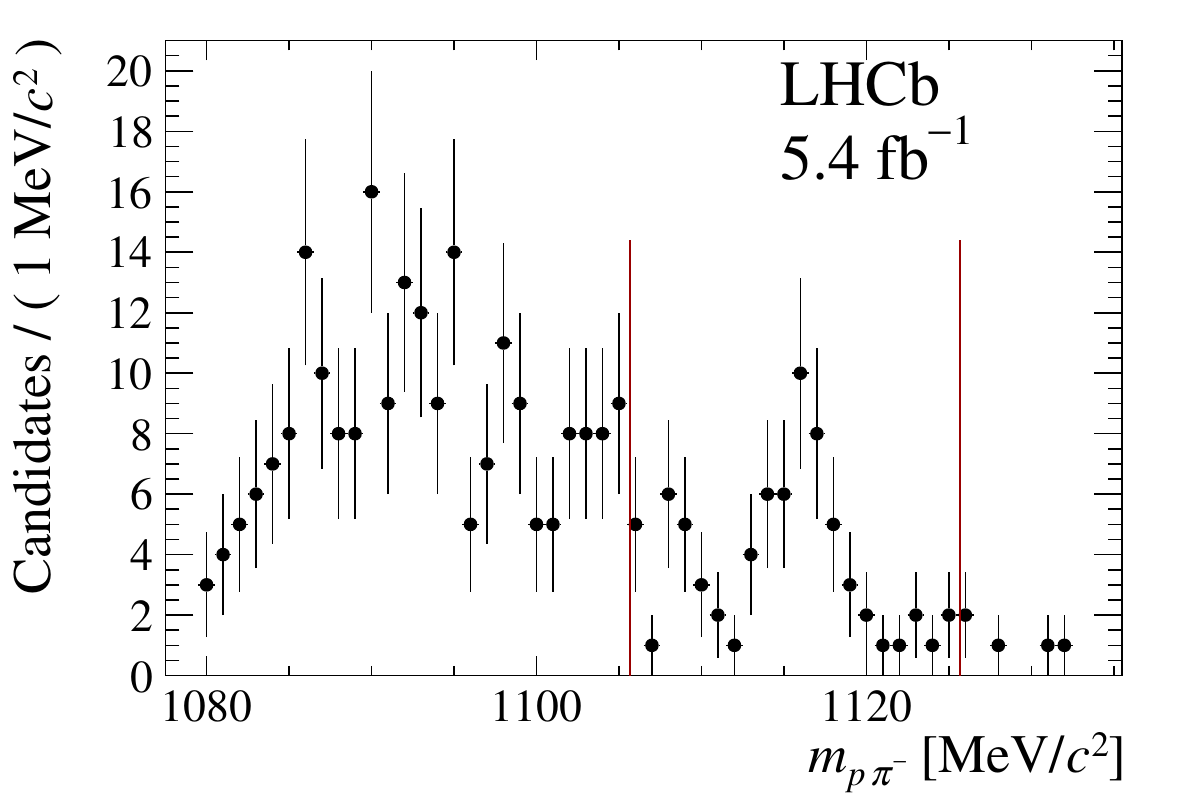}
    \caption{Distribution of the $p\mu^-$ invariant mass in the $p\pi^-$ hypothesis for \sigmapmumu  candidates in data 
    within a $\pm 6 \mevcc$ window from the $\Sigmap$ mass peak, without the \Lz veto (indicated between red lines).}
    \label{fig:lambdanoveto}
\end{figure}

The distribution of the $p \mu^-$ invariant mass in the $p\pi^-$ hypothesis is reported in Fig.~\ref{fig:lambdanoveto} for \sigmapmumu candidates in data, without the \Lz veto (shown as red lines). 

\begin{figure}[!h]
    \centering
    \includegraphics[width = 0.48\textwidth]{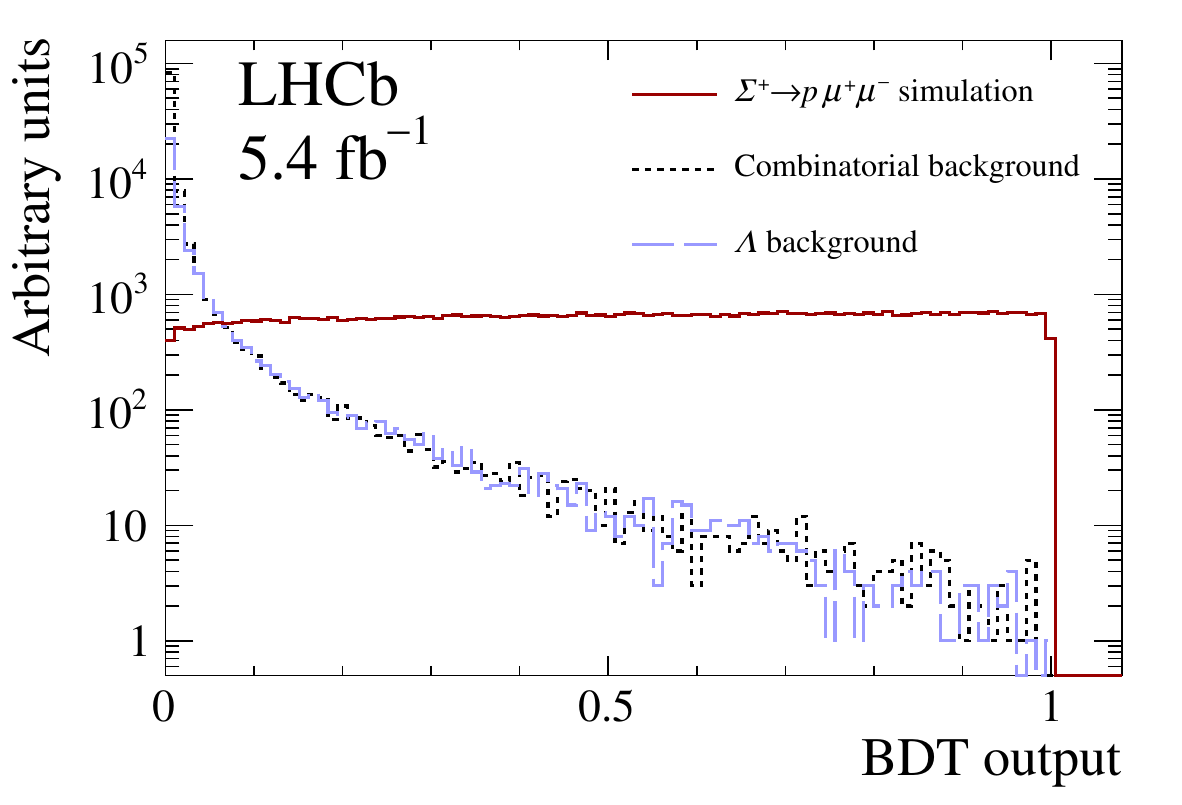}
        \includegraphics[width=0.48\textwidth]{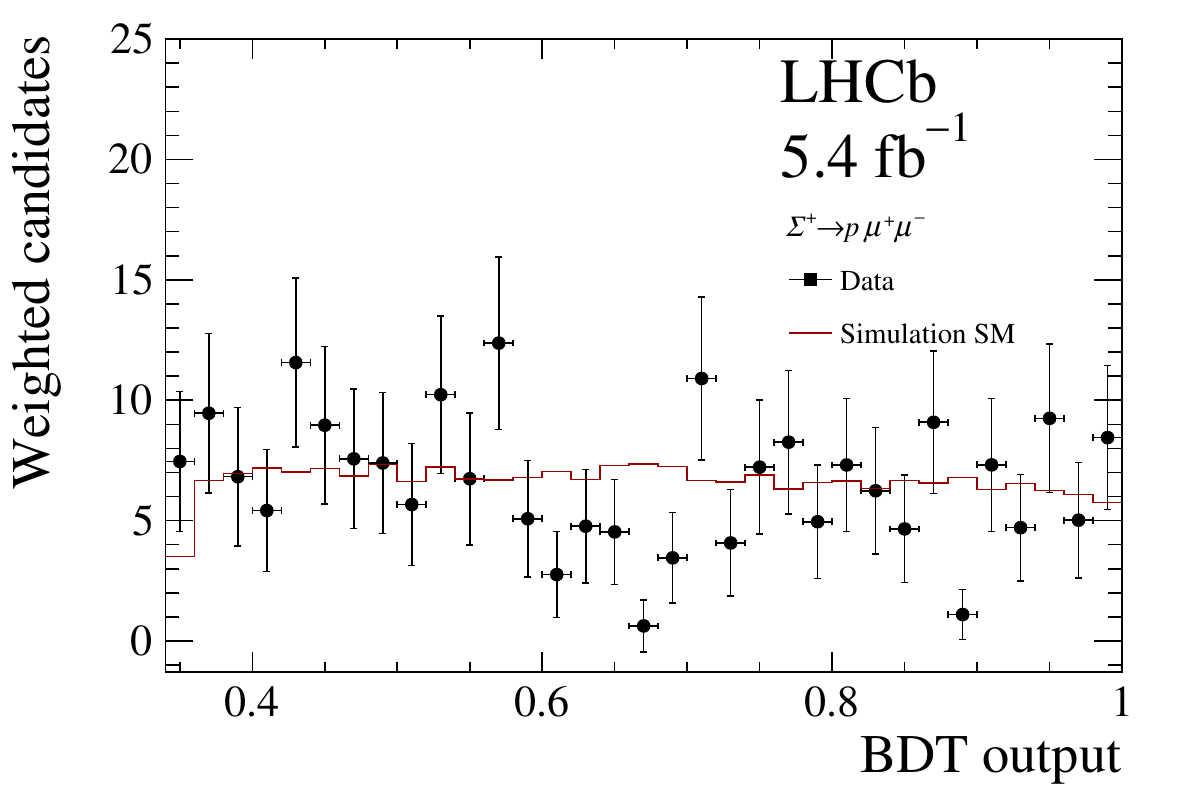}
    \caption{Distribution of the BDT output variable for (left) simulated signal (red solid line) and data divided into combinatorial background (black dashed line) and $\Lz$ background (blue long-dashed line), and (right)  for background subtracted \sigmapmumu decays in data (black dots) and in simulation (red line). The uncertainty on the SM weight in the simulation is not shown.}
    \label{fig:bdtdistro}
\end{figure}

The distribution of the BDT output is shown in Fig.~\ref{fig:bdtdistro}(left) for simulated signal events and for background in data from the sidebands of the invariant-mass distribution (candidates in the range $\mpmumu < 1173\mevcc$ or $1205<\mpmumu<1400\mevcc$). Candidates in data are divided into those that pass the \Lz veto and those that do not pass it.
Figure~\ref{fig:bdtdistro}(right) shows the BDT output distribution for candidates in data above the chosen threshold value and background subtracted, compared with the signal simulated sample. 

\begin{figure}[!btp]
\centering
\includegraphics[width=0.48\textwidth]{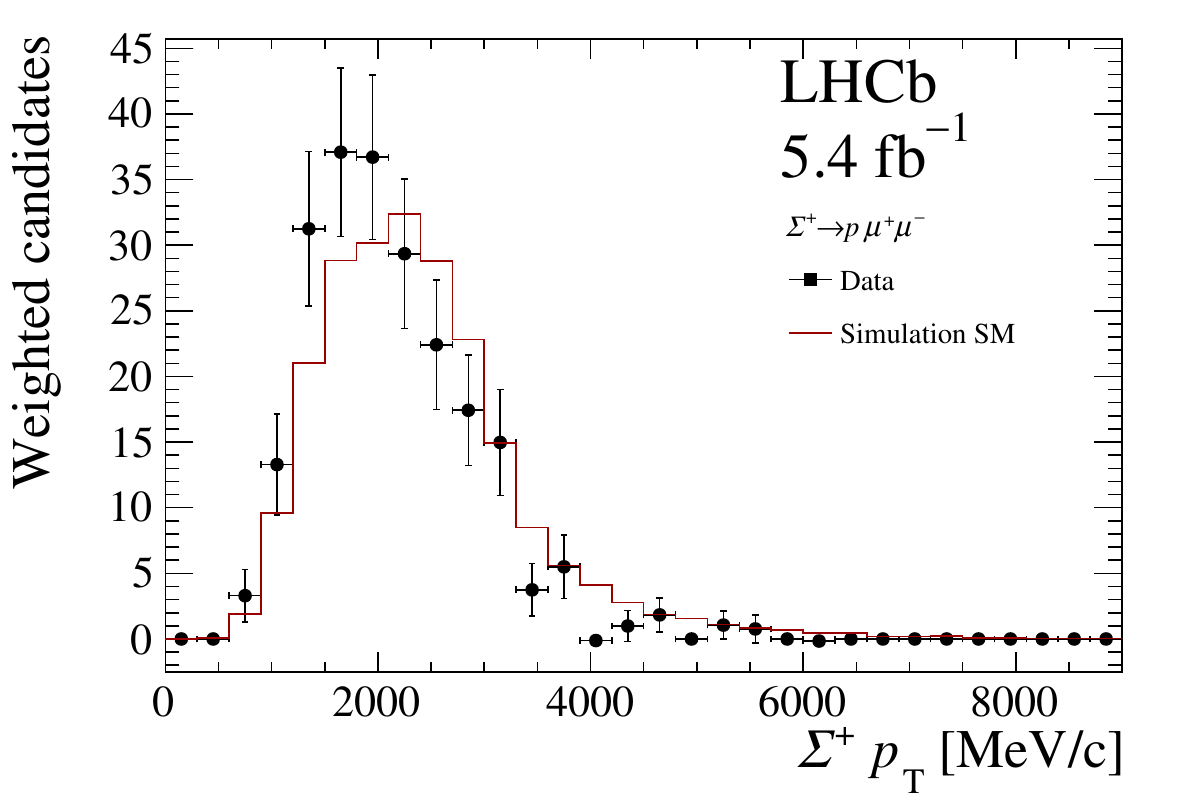}
\includegraphics[width=0.48\textwidth]{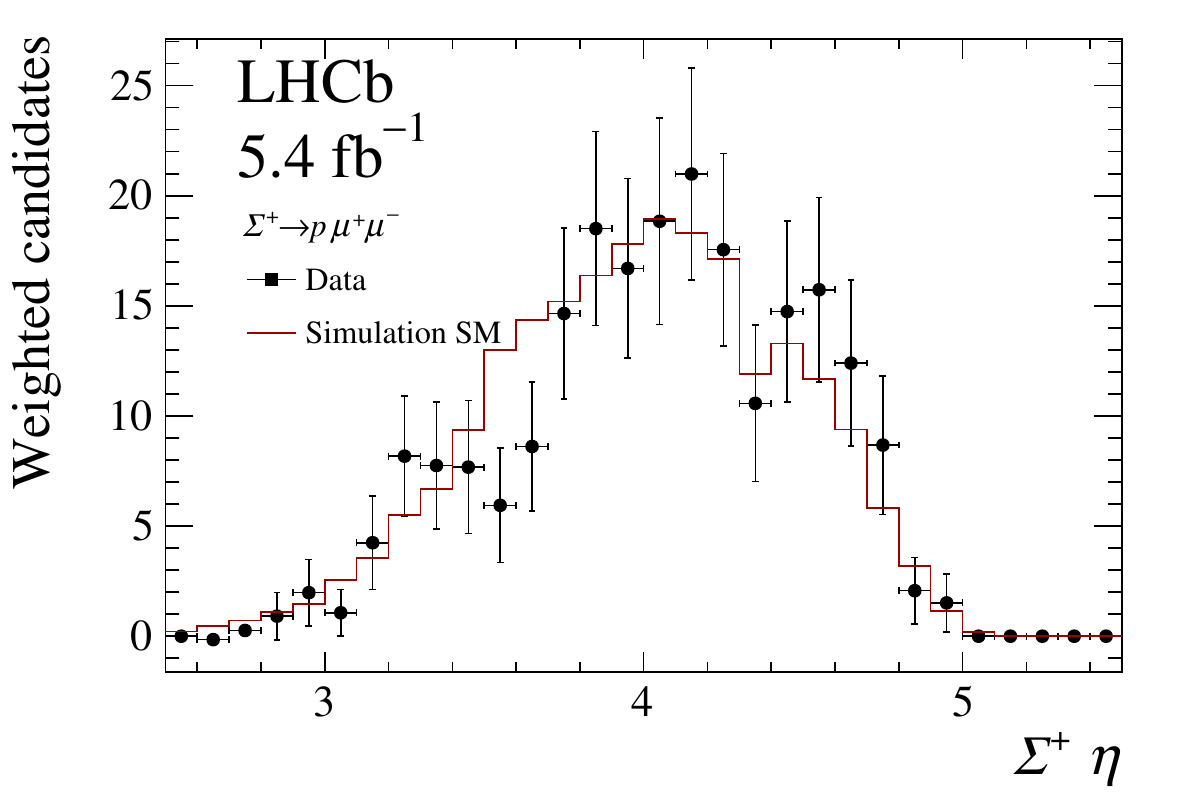}
\caption{ Distribution of (left) the \pt and (right) pseudorapidity ($\eta$) variables for background subtracted \sigmapmumu decays in data (black dots) and in simulation (red line). 
The uncertainty on the SM weight in the simulation is not shown. }
\label{fig:sigmakinematics}
\end{figure}

\begin{figure}[h]
\centering
\includegraphics[width=0.48\textwidth]{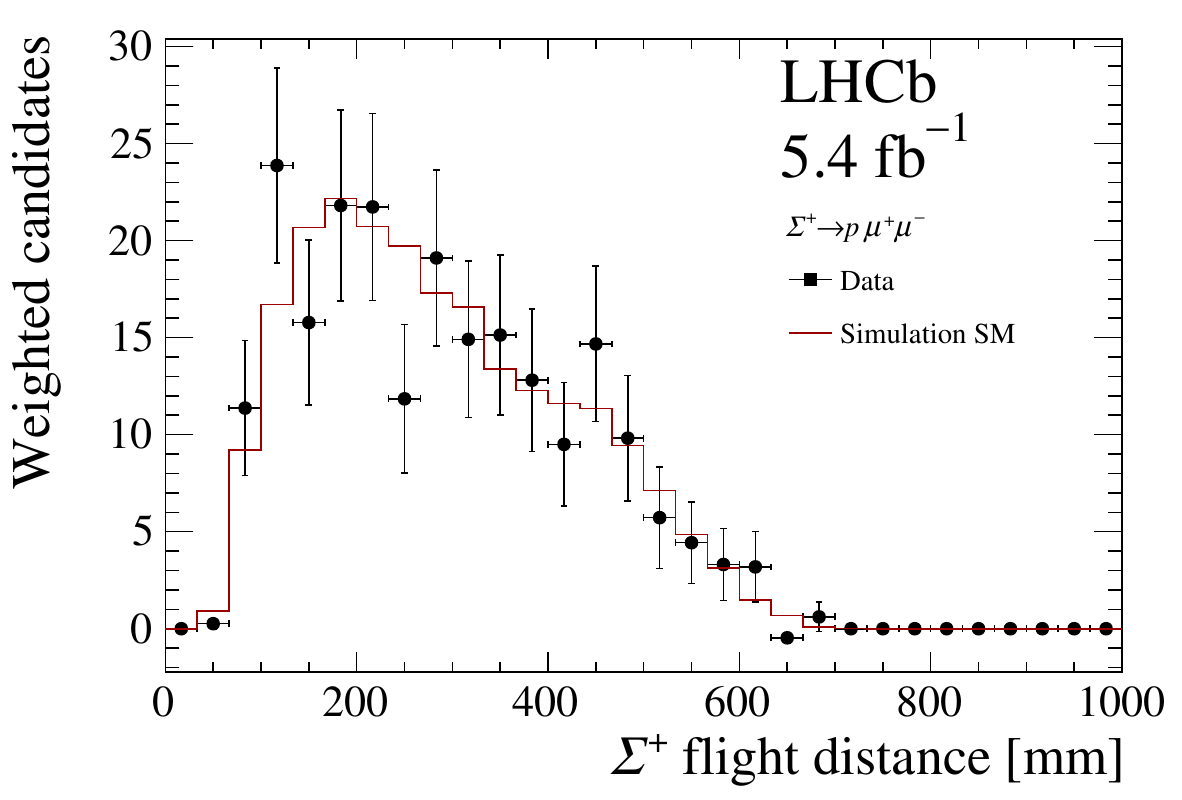}
\caption{ Distribution of the $\Sigma^+$ flight distance from the primary vertex  for background subtracted \sigmapmumu decays in data (black dots) and in simulation (red line). The uncertainty on the SM weight in the simulation is not shown. }
\label{fig:sigmafd}
\end{figure}

Figures~\ref{fig:sigmakinematics} and \ref{fig:sigmafd} show the distribution in background-subtracted \sigmapmumu candidates in data for the \Sigmap transverse momentum (\pt), pseudorapidity ($\eta$) and flight distance with respect to the primary vertex. 
The distributions are compared with the ones from the signal simulated sample. 

\clearpage


\addcontentsline{toc}{section}{References}
\bibliographystyle{LHCb}
\bibliography{main,standard,LHCb-PAPER,LHCb-CONF,LHCb-DP,LHCb-TDR}

\newpage
\centerline
{\large\bf LHCb collaboration}
\begin
{flushleft}
\small
R.~Aaij$^{38}$\lhcborcid{0000-0003-0533-1952},
A.S.W.~Abdelmotteleb$^{57}$\lhcborcid{0000-0001-7905-0542},
C.~Abellan~Beteta$^{51}$\lhcborcid{0009-0009-0869-6798},
F.~Abudin{\'e}n$^{57}$\lhcborcid{0000-0002-6737-3528},
T.~Ackernley$^{61}$\lhcborcid{0000-0002-5951-3498},
A. A. ~Adefisoye$^{69}$\lhcborcid{0000-0003-2448-1550},
B.~Adeva$^{47}$\lhcborcid{0000-0001-9756-3712},
M.~Adinolfi$^{55}$\lhcborcid{0000-0002-1326-1264},
P.~Adlarson$^{84}$\lhcborcid{0000-0001-6280-3851},
C.~Agapopoulou$^{14}$\lhcborcid{0000-0002-2368-0147}
% ...

{\footnotesize \it

% ...
$^{14}$Universit{\'e} Paris-Saclay, CNRS/IN2P3, IJCLab, Orsay, France\\
% ...
$^{38}$Nikhef National Institute for Subatomic Physics, Amsterdam, Netherlands\\
% ...
$^{47}$Instituto Galego de F{\'\i}sica de Altas Enerx{\'\i}as (IGFAE), Universidade de Santiago de Compostela, Santiago de Compostela, Spain\\
% ...
$^{51}$Physik-Institut, Universit{\"a}t Z{\"u}rich, Z{\"u}rich, Switzerland\\
% ...
$^{55}$H.H. Wills Physics Laboratory, University of Bristol, Bristol, United Kingdom\\
% ...
$^{57}$Department of Physics, University of Warwick, Coventry, United Kingdom\\
% ...
$^{60}$School of Physics and Astronomy, University of Glasgow, Glasgow, United Kingdom\\
$^{61}$Oliver Lodge Laboratory, University of Liverpool, Liverpool, United Kingdom\\
% ...
$^{69}$Syracuse University, Syracuse, NY, United States\\
% ...
$^{84}$Department of Physics and Astronomy, Uppsala University, Uppsala, Sweden, associated to $^{60}$\\
% ...
}
\end{flushleft}

\end{document}